\documentstyle[12pt]{article} 
 
\newcommand{\uD}{\mbox{I} \! \mbox{D}}
\newcommand{\uE}{\mbox{I} \! \mbox{E}}
\newcommand{\uG}{\mbox{G} \! \! \! \! \mbox{\sf I} \: \:}
\newcommand{\uI}{\mbox{I} \! \mbox{I}}
\newcommand{\uM}{\mbox{I} \! \mbox{M}}
\newcommand{\uR}{\mbox{I} \! \mbox{R}}
\newcommand{\E}{{\scriptstyle\mbox{\scriptsize I}\!\mbox{\scriptsize E}}} 
\newcommand{\vtD}{{\uD}}
\newcommand{\vtE}{{\uE}}
\newcommand{\ivtE}{{\E}} 
\newcommand{\vtG}{{\uG}}
\newcommand{\vtH}{{H}} 
\newcommand{\vtHH}{\vtH_{\vc,\vA}}
\newcommand{\vtI}{{\uI}}
\newcommand{\vtM}{{\uM}}
\newcommand{\vtR}{{\uR}}
\newcommand{\vtV}{{V}} 
\newcommand{\atE}{{E}}
\newcommand{\atI}{{I}}
\newcommand{\atM}{{M}}
\newcommand{\vF}{{F}} 
\newcommand{\vW}{{W}} 
\newcommand{\vpi}{\mbox{\boldmath $\pi$}} 
\newcommand{\vtau}{\mbox{\boldmath $\tau$}} 
\newcommand{\vb}{{b}} 
\newcommand{\vc}{{a}} 
\newcommand{\vg}{{g}} 
\newcommand{\vj}{{\eta}} 
\newcommand{\vk}{{k}} 
\newcommand{\vq}{{q}} 
\newcommand{\vr}{{r}} 
\newcommand{\vt}{{t}} 
\newcommand{\vu}{{u}} 
\newcommand{\vv}{{v}} 
 
\newcommand{\vx}{{x}} 
\newcommand{\vA}{{A}}

\newcommand{\br}{\mbox{\boldmath $r$}} 
\newcommand{\bv}{\mbox{\boldmath $v$}}

\newcommand{\sW}{{\scriptscriptstyle W}}   

\newcommand{\KK}{K_{\vc,t}}
\newcommand{\VV}{V_{\vc,t}}
\newcommand{\cE}{{\cal E}} 
\newcommand{\cS}{{\cal S}} 
\newcommand{\tF}{{\tilde F}} 
 
\newcommand{\sa}{\Gamma} 
\newcommand{\ray}{\Theta}
\newcommand{\dl}{l}
\newcommand{\ep}{\varepsilon} 
\newcommand{\ro}{\varrho}
\newcommand{\2}{{(2)}}               
\newcommand{\gyok}{^{1/2}} 
\newcommand{\fel}{(1/2)}
\newcommand{\fsize}{\footnotesize}
\newcommand{\fegy}{\mbox{\fsize 1}}
\newcommand{\fketto}{\mbox{\fsize 2}}
\newcommand{\fnegy}{\mbox{\fsize 4}}
\newcommand{\fe}{\mbox{\fsize {\it e}}}
\newcommand{\fimag}{\mbox{\fsize {\it i}}}
\newcommand{\fk}{\mbox{\fsize {\it k}}}
\newcommand{\fm}{\mbox{\fsize {\it m}}}
\newcommand{\fketm}{\mbox{\fsize 2\hspace{-.1ex}{\it m}}}

\newcommand{\fketim}{\mbox{\fsize 2\hspace{-.07ex}{\it i\hspace{.04ex}m}}}
\newcommand{\fhbara}{{\it h}\hspace{-.97ex}\rule[1.3ex]{.76ex}{.08ex}}
\newcommand{\fhbarb}{\fhbara\hspace{.21ex}}
\newcommand{\fhbarc}{\fsize {\fhbarb\hspace{.14ex}}}
\newcommand{\fhbar}{\mbox{\fsize \fhbarb}}
\newcommand{\fmnegyzet}{\mbox{\fm\hspace{.1ex}\rbox{.62ex}{\tiny 2}}}
\newcommand{\fketmnegyzet}{\mbox{\fketto\hspace{-.1ex}\fmnegyzet}}
\newcommand{\fnegymnegyzet}{\mbox{\fnegy\hspace{-.1ex}\fmnegyzet}}
\newcommand{\fhbarnegyzet}{\mbox{\fhbarc\rbox{1.1ex}{\tiny 2}}}
\newcommand{\fspace}{\hspace{0.25ex}}
\newcommand{\rbox}[2]{\raisebox{#1}{#2}}
\newcommand{\ftort}[2]{\frac{\rbox{-.3ex}{#1}}{\rbox{.48ex}{#2}}}
 
\newcommand{\Laplace}{\bigtriangleup}
\newcommand{\parc}{\partial} 
\newcommand{\aranyos}{\sim}
\newcommand{\unio}{\cup}
\newcommand{\metszet}{\cap}
\newcommand{\kb}{\approx}          
\newcommand{\ortog}{\perp}         
\newcommand{\minden}{\forall}      
\newcommand{\barmely}{\forall}     
\newcommand{\tenzor}{\otimes}      
\newcommand{\ek}{\wedge}           
\newcommand{\es}{\wedge}           
\newcommand{\vagy}{\vee}           
\newcommand{\kor}{\circ}           
\newcommand{\pont}{\dot{}}         
\newcommand{\x}{\times}            
\newcommand{\azonos}{\equiv}       
 
\newcommand{\ar}{\left(\begin{array}{c}} 
\newcommand{\arr}{\left(\begin{array}{cc}} 
\newcommand{\arrr}{\left(\begin{array}{ccc}} 
\newcommand{\ay}{\end{array}\right)} 
 
\newcommand{\point}{$\!\!\!\!\!\!\!\!\!$ . $\,$} 
\newcommand{\nn}{\nonumber \\} 
\newcommand{\ea}{\end{eqnarray}} 
\newcommand{\ee}{\end{equation}} 

\def\_#1{^{}_{#1}}                 
\def\ba#1{\begin{eqnarray}\label{eq:#1}} 
\def\be#1{\begin{equation}\label{eq:#1}} 
\def\l#1{\label{eq:#1}} 
\def\r#1{(\ref{eq:#1})} 
\begin{document} 
 
\title{
\mbox{}\vspace*{-3.0cm} \\
{\bf Building a frame and gauge free formulation of quantum mechanics}}
\author{T. F\"ul\"op\thanks{On leave from Institute for  
   Theoretical Physics, E\"otv\"os University, Budapest, Hungary} 
   \thanks{Electronic mail: fulopt@tanashi.kek.jp} \\  
   {\it Institute of Particle and Nuclear Studies,} \\ 
   {\it High Energy Accelerator Research Organization (KEK)} \\
   {\it Tanashi Branch;}
   {\it Midori, Tanashi, Tokyo 188-8501, Japan} 
   \and S. D. Katz\thanks{Electronic mail: katz@ludens.elte.hu} \\ 
   {\it  E\"otv\"os University, Budapest} \\
   {\it M\'uzeum krt. 6-8, Budapest 1088, Hungary}}
\date{}
\maketitle

   \begin{abstract} 
 
The wave function of quantum mechanics is not a boost invariant and gauge
invariant quantity. Correspondingly, reference frame dependence and gauge
dependence are inherited to most of the elements of the usual formulation
of quantum mechanics (including operators, states and events). If a frame
dependent and gauge dependent formalism is called, in short, a relative 
formalism, then the aim of the paper is to establish an absolute, i.e.,
frame and gauge free, reformulation of
quantum mechanics. To fulfil this aim, we develop absolute quantities and
the corresponding equations instead of the wave function and the
Schr\"odinger equation. The absolute quantities have a more direct
physical interpretation than the wave function has, and the corresponding
equations express explicitly the independent physical aspects of the
system which are contained in the Schr\"odinger equation in a mixed and
more hidden form. Based on the absolute quantities and equations, events,
states and physical quantities are introduced also in an absolute way. 
The formalism makes it possible to obtain some sharper versions of the
uncertainty relation and to extend the validity of Ehrenfest's theorem. 
The absolute formulation allows wide extensions of quantum mechanics. To
give examples, we discuss two known nonlinear extensions and, in close
details, a dissipative system. An argument is provided that the absolute
formalism may lead to an explanation of the Aharonov-Bohm effect
purely in terms of the electromagnetic field strength tensor. At last, on
special relativistic and curved spacetimes absolute quantities and
equations instead of the Klein-Gordon wave function and equation are
given, and their nonrelativistic limit is derived. 

   \end{abstract}


   \section{\point Introduction}


Quantum mechanics, being an amazingly successful, and, at the same time, a
surprisingly novel theoretical framework in theoretical physics, has been
investigated since its birth from innumerable aspects. As time passes, our
understanding of it becomes wider and wider by the new properties
explored. However, still there exist numerous questions of various kind
concerning quantum mechanics that need further thorough investigation,
directions in which our present understanding is not satisfactory yet. 
This paper is devoted to one such question, a topic that arises about the
formulation and has diverse physical motivations. 

The starting observation is that the usual formalism of quantum
mechanics---wave functions, Hilbert space operators, Schr\"odinger
equation, etc.---, no matter the background spacetime is
nonrelativistic, special relativistic or a curved one, needs a choice of a
reference frame on the spacetime in question. Based on a reference frame,
spacetime is split into space and time, and the Schr\"odinger equation
describes the time evolution of the wave function, a square integrable
function of the space variable. Events, physical quantities and states
(density operators) are introduced corresponding to the Hilbert space of
square integrable functions. If another reference frame is chosen on the
spacetime, all these quantities need to be transformed, they are not
invariant under the transformation of the reference frame. In the
nonrelativistic case this property is transparent, the wave function is
transformed by a spacetime depending multiplying function, and the
transformation of all the other quantities correspond to the
transformation of the wave function. In the case of the Klein-Gordon
equation this frame dependence is not so apparent since this equation has 
a ``covariant'' form; however, the way in which 
physical quantities can be introduced and in which the physical 
interpretation of
the Klein-Gordon theory can be given requires a choice of a reference
frame (see \cite{FesVil} for special relativistic spacetime and \cite{FT2}
for static curved spacetimes) so the full formalism proves to be frame
dependent indeed. 

In the case of nonrelativistic or special relativistic spacetime this
feature does not have serious consequences because of the equivalence of
(inertial) reference frames. However, on a curved spacetime it leads to a
physically problematic situation both in quantum mechanics and in quantum
field theory: On one hand, different choices of reference frames lead to
different quantizations resulting physically inequivalent quantum systems,
and, on the other hand, the time variable introduced this way generally
cannot be regarded a physical one (the elapsed proper time between two
spacelike hypersurfaces is space dependent in general). In quantum
gravity, the corresponding situation is called ``the problem of time''
\cite{ASpot,Kuc,Ish}. 

Another observation about the formalism of quantum mechanics is that the
wave function, and, correspondingly, operators, events and states, are
gauge dependent quantities. More closely, when a charged particle is
considered in a given electromagnetic field, the Hamiltonian of the system
is constructed from the electromagnetic four-potential. This leads to the
consequence that different choices of the potential lead to a spacetime
dependent transformation of the wave function. The formalism of quantum
field theory inherits this property from quantum mechanics. 

In addition, the wave function possesses another, smaller but also
important, arbitrariness, an ambiguity up to an arbitrary constant phase
factor. It is in fact a ray in the Hilbert space, rather than an element
of the Hilbert space, that bears a physical meaning.

If we call a frame dependent and/or gauge dependent formalism a relative
formalism, then the aim of the approach presented in this paper can be
expressed as to develop an absolute formalism for quantum mechanics. A
frame free and more spacetime friendly formalism can shed new light on, and
can give a better understanding of, the known troublesome properties of
quantum theories on curved spacetimes. In parallel, a gauge independent
formalism for quantum mechanics, and for quantum field theory, may have
natural advantages, too (see, e.g.,\ \cite{KT} and references therein). 
Furthermore, by their very nature, absolute quantities have a more direct
physical interpretation, thus they are interesting for measurement theory
as well. These are the initial motivations to seek an absolute formalism.
On the other hand, it will turn out that, as we start to build up our
approach---in this paper we will concentrate basically on (one-particle
zero spin) nonrelativistic quantum mechanics---, new additional benefits
will appear immediately. 

Spacetime is a given background in quantum mechanics, and an absolute
formulation of quantum mechanics needs first an absolute spacetime
formalism. The first appearance of such a formalism can be found in
\cite{Weyl}. Since then several similar treatments have been worked out, 
and,
for example, in general relativity and manifold theory the coordinate free
language is widely used. For nonrelativistic and special relativistic
spacetime a profound and detailed presentation can be found in \cite{MT3}
(see also \cite{MT1}), 
and in our approach we will use the treatment appearing
there. The basic features of nonrelativistic spacetime are the affine
structure of spacetime, an absolute time structure on it, and a Euclidean
structure on the hyperplanes of simultaneous spacetime points. It is
important to note that space is not absolute even nonrelativistically,
different reference frames have different own spaces---for example, the
formula of the special Galilean transformation rule, $ t' = t $, $ \br' =
\br - \bv t $ (with the usual notations), reflects this in such a way
that under a transformation `position is not mixed into time' (time is
absolute) but `time is mixed into position' (space is relative). 

Next, an absolute formalism for classical mechanics is needed. To build up
an absolute presentation of classical mechanics, world lines and the
Newton equation can be formulated in an absolute way in a straightforward
manner. On the other side, as space is frame dependent, and as, for
conservative forces (such as the electromagnetic Lorentz force), canonical
momenta are both frame and gauge dependent, phase space is not an absolute
concept. As observed by Souriau \cite{Sou} (see also \cite{MT2}), it is
the space of the solutions of the equation of motion (in short,
processes) that is the corresponding absolute object. In the case of
conservative forces this process space admits an absolute symplectic
structure, as an absolute counterpart of the symplectic structure of the
phase space. Events, states and physical quantities (observables), whose 
usual introduction is based on the phase space, can be given an absolute 
definition, too, on the basis of the process space \cite{MT2}. 

For quantum mechanics, a similar approach is developed in \cite{MT2}.
There the Schr\"odinger equation is understood as a partial differential
equation for complex functions on spacetime, processes are the solutions
of the Schr\"odinger equation, and the process space is found to admit a
Hilbert space structure. Then, events, states and physical quantities are
introduced with respect to the Hilbert space of {\sl processes}. 

Unfortunately, that formalism of quantum mechanics eliminates
only a part of relativeness. The basic problem is the fact
that the Schr\"odinger equation, even in its most absolute form, needs a
choice of a gauge and a four-velocity value (the latter is mathematically
equivalent to a choice of an inertial observer). Hence, to obtain an
absolute approach, here our strategy is, first, to find such new
quantities and equations instead of the wave function and the
Schr\"odinger equation that do not require these choices, and second, to 
build from the resulting absolute process space the event space also in an
absolute way, and third, to introduce states and physical quantities
based on the absolute event space. Thus the whole theory is given in an
absolute form. 

Concerning the first step, our starting point will be to make use of the
polar decomposition of the wave function. This method of introducing gauge
independent (and ray ambiguity free) quantities and equations that are
equivalent to the wave function and the Schr\"odinger equation has already
been applied by several authors \cite{Mad,Tak,Jan,BH,KR}. 
Since here our further requirement is to be frame free, too, we will
construct such new quantities and equations that fulfil all these demands.
Instead of the complex valued wave function, our absolute process function
turns out to be a real cotensor field. (Actually, to offer flexibility, we
will provide different equivalent sets of absolute quantities and
equations to describe the absolute processes in different ways.) The 
obtained 
absolute equations exhibit the three independent physical aspects of the
system. In the case of the Schr\"odinger equation these aspects are
present in a mixed and more hidden form. 

The geometry of the resulting absolute process space proves to be
different from the geometry of a Hilbert space. The structure of the
`intermediate step', the space of rays of a Hilbert space---which is
studied in \cite{AS}---stands closer to it. Namely, it will turn out that
the absolute process space is a complex Hilbert manifold. As will be
useful to observe, an equivalent description is also available: It can be
considered as a real K\"ahler manifold, too, including a Riemannian and a
symplectic structure. This latter aspect is useful, for example, for
expressing the geodesic curves of the absolute process space in a 
transparent way. 

The next step is to derive the event space from the absolute process
space, based on the explored properties of the latter. The events of
quantum mechanics, which in a Hilbert space formalism are linear subspaces
of the Hilbert space, are formulated here as appropriate subsets of the
absolute process space. The structures of the quantum mechanical event
space, including the operations `and', `or' and `negation', are also given 
an absolute definition. 

Then states and physical quantities can be defined with respect to this
event space formulation. To give examples, some physical
quantities---position, velocity, kinetic energy---will be discussed more
closely. As a part, we show, by means of the absolute quantities, a simple
derivation of the position-velocity uncertainty formula. (We prefer
velocity to canonical momentum since velocity is a gauge free quantity
while canonical momentum is not.) Indeed, we will obtain some sharper
inequalities than the usual uncertainty formula. The Ehrenfest theorem
will be presented, too, by the use of absolute quantities and equations.
It will be a remarkable observation that this theorem proves to be valid
not only for conservative forces (e.g.,\ the Lorentz force) but for more
general ones, too. 

Actually, we will see that the whole absolute formalism accepts more
general types of forces than conservative ones. One of the opening
possibilities is the treatment of dissipative forces. Another opportunity
is to use the absolute framework in the search of generalizations of
quantum mechanics. The latter topic is also mentioned as the search for
nonlinear extensions of the Schr\"odinger equation. A well-known such
extension is the so-called nonlinear Schr\"odinger equation, and other
ones are worked out and discussed in \cite{BBM} and in \cite{W}, for
example. A framework for generalizations of quantum mechanics is given by
\cite{AS}, in a ray ambiguity free treatment. The absolute formalism
provides an approach which, in addition, is frame free and gauge free. 
Thus it gives a convenient and safe framework to introduce new
generalizations of quantum mechanics, or to test already existing ones. 
We will demonstrate this by checking that both the so-called nonlinear
Schr\"odinger equation and the equation proposed by Byalinicki-Birula and
Mycielski \cite{BBM} satisfy the requirement of having a ray ambiguity
free, frame free, and gauge free physical content. 

Concerning the treatment of dissipative forces, it is important to note
that, in contrast to a Schr\"odinger equation with a nonhermitian
Hamiltonian, the dynamics given by the absolute equations of a 
dissipative system is `unitary', i.e., the total probability
of finding the particle is time independent. It is the mechanical
energy that decreases in time. These properties are in accord with the 
physical expectations. 

We will investigate a dissipative example, the case of the damping force
with linear velocity dependence, in detail. We will be interested in the
time dependence of the expectation value and uncertainty of position and 
of the expectation value of kinetic energy. In
particular, an interesting question is whether this damping force stops
the spread of a `wave packet' that is unavoidably present in the free
particle case, whether the `wave packet' (the process function) will tend
to a stationary solution. To answer these questions (for simplicity, the
calculations are done for the $ 1 + 1 $ dimensional system), first we 
determine the
stationary solutions of the equations. It turns out that there is no
normalizable stationary solution, hence there is no limiting stationary
solution the process function could tend to. Then, for arbitrary initial
conditions, we determine the exact time dependence of the expectation
value of position, and the asymptotic time dependence of the
uncertainty of position and of the expectation value of kinetic
energy. We find that the expectation value of position stops
exponentially. Furthermore, the spread of the `wave packet' is found to be
slower than in the free particle case---the asymptotic time dependence of 
the uncertainty
of position is proportional to $t^{1/4}$ instead of $t$---but, as the
result shows, this spread will never stop. In parallel, the
expectation value of kinetic energy dissipates completely, it tends to
zero, following a $t^{-1/2}$ asymptotic time dependence. 

Turning back to the case of the charged particle in an electromagnetic
field, there the absolute formalism works with the electromagnetic
field strength tensor instead of a corresponding four-potential. This is
an interesting feature from the aspect of the Aharonov-Bohm effect (see
\cite{AB} for the necessary details). In this effect the particle is
excluded from a cylindrical space region, however, its motion is
influenced by a magnetic field which is nonzero only within this
region.The quantitative apperance of the Aharonov-Bohm effect is the
integral of the vector potential along a closed curve around the cylinder
which hides the magnetic field. This quantity is just the flux of the
magnetic field within the cylinder, thus it can be expressed by means of
the electromagnetic field strength tensor only. However, if we think in
terms of the field strength, it is physically strange that the particle is
excluded from the inside of the cylinder by a high potential barrier,
nevertheless, it `observes' the magnetic field which is present only 
inside the cylinder.
It turns out that the absolute formalism seems to be promising to
understand this effect only by means of the electromagnetic field strength
tensor. Namely, we will show a model computation which gives the idea that
while the position probability density is excluded from the inside of the
cylinder, there are other absolute quantities---being also necessary for
the complete description of the particle---which do not vanish within the
cylinder. They seem to penetrate unavoidably into the cylinder, which
suggests the picture that these quantities are those through which the
particle `feels' the inner magnetic field. 

Concerning (zero spin) relativistic quantum mechanics, in this paper we
take the first step: To give absolute quantities and equations equivalent
to the Klein-Gordon process function and the Klein-Gordon equation. We do
it both on special and general relativistic spacetime. Then in the special
relativistic case we show how the nonrelativistic limit of the
quantities and equations can be performed. Similarly, in the case of 
curved spacetime we
carry out the nonrelativistic plus weak gravity limit of the quantities
and equations. After finding the appropriate way to deduce the
nonrelativistic quantities from the general relativistic ones, the result
will be the expected one: The appearance of the nonrelativistic form of
the gravitational force in addition to the effect of the electromagnetic
field. 

The paper is organized as follows. The elements of the absolute
description of nonrelativistic spacetime necessary for our considerations
are collected and summarized in Sect.~II\@. In Sect.~III. the absolute
equivalents of the wave function and the Schr\"odinger equation are
given. We explore the geometric properties of the absolute process space
in Sect.~IV\@. Events, states and physical quantities are introduced and
discussed in Sect.~V. and Sect.~VI\@. Sect.~VII. deals with the
`nonlinear' and the dissipative generalizations of quantum mechanics, and
Sect.~VIII. presents the calculations about the dissipative example
system mentioned above. The Aharonov-Bohm effect is investigated in
Sect.~IX\@. The relativistic quantities and equations, and their limiting
cases are shown in Sect.~X\@. Sect.~XI. contains the Discussion and
outlook. 

Finally we remark that we will use a formalism which does not need 
choices of measurement units when dealing with physical quantities 
\cite{MT3,MT1}. For example, distance values, time period values and 
mass values are physically different. To reflect this, instead of using 
the line of real numbers, $\vtR$, for all these kinds of quantities, we 
will use different one dimensional vector spaces---called measure 
lines---for them. Products and quotients of quantities of different 
dimensions are established by tensorial products and quotients of one 
dimensional vector spaces, and a choice of a measurement unit is 
formulated as a choice 
of an element of the measure line---the choice of a basis in the one 
dimensional vector space, indeed. As one can check, this treatment 
reflects exactly the expected properties and rules concerning how to 
handle dimensionful quantities, for example, the ones concerning 
multiplication of quantities of different dimensions, or the ones about 
changing measurement units.  
 
Actually, a general one dimensional vector space differs from $\vtR$ only
in that in $\vtR$ there is a distinguished unit element, otherwise all the
linear properties are exactly the same. Thus, while being technically
easy, this use of one dimensional vector spaces ensures a correct handling
(`bookkeeping') of the different dimensions and makes it possible to work
with physical quantities without choosing measurement units. Actually, a
choice of a measurement unit is also a relative step, not belonging to the
examined physical system but to the way we are doing our observations
about it. The exception is when the physical system or phenomenon itself
owns some dimensionful parameters. Then these values offer distinguished
identifications between different measure lines. For instance, in
relativistic physics the speed of light offers an identification between
space distances and time intervals. In some cases there are enough
dimensionful parameters that all quantities can be made
dimensionless---the dissipative system treated in detail in our
Sect.~VIII. will be such an example. 
 
The measure lines of distance values, time intervals and mass values will
be denoted by $\vtD$, $\vtI$, and $\vtG$, respectively. As, for instance,
$ \vtD^* \azonos \vtR / \vtD $, we see that inverse length values, e.g.,\
$ 1 cm^{-1} $, `reside' in $\vtD^*$. As another example, a magnitude of a 
force is an element of $ \vtG \tenzor \vtD / \vtI^\2 = \vtG \tenzor \vtD
\tenzor (\vtI^*)^\2 $ (notation: $ \vtI^\2 = \vtI \tenzor \vtI $).


   \section{\point The nonrelativistic spacetime model}


A useful way to collect and present the necessary properties of
nonrelativistic spacetime is to do it through a comparison with the
corresponding elements of special relativistic spacetime. In this way the
basic similarities and differences are well illuminated. Therefore, first 
we give a short overview of the special relativistic spacetime. Both 
spacetimes will be treated in the absolute formalism of \cite{MT3} (see 
also \cite{MT1}). 

A general relativistic, or curved, spacetime is given by $\atM$, $\vtI$,
and $\vg$, where $\atM$ is a four dimensional oriented manifold, $\vtI$ is
the measure line of time intervals, and $\vg$ is an arrow oriented Lorentz
form on $\atM$. Special relativistic spacetime is the special case when
$\atM$ is an affine space and $\vg$ is a constant tensor field on it. In
this case the tangent bundle is of the form $ \atM \x \vtM $, the tangent
space at each spacetime point is $\vtM$, where $\vtM$ is the underlying
vector space of $\atM$. World lines, which are the models for pointlike 
classical material objects, are curves on spacetime with timelike tangent 
vectors 
at each point. Between two points of a world line the proper time length
(taking its value in $\vtI$) can be defined with the aid of $\vg$. Usually
world lines are considered with a proper time parametrization on them. The
tangent vectors of a world line obtained by derivation with respect to
proper time are elements of $ \vtM / \vtI $ and satisfy the following
relation: 
   \be{V1} 
   \vg(\vu,\vu) \azonos \vu \cdot \vu = 1 
   \ee
 [our metric convention is $(+,-,-,-)$; inner products, and the
action and composition of linear maps, are denoted by a dot product]. 
The futurelike elements of $ \vtM / \vtI $ satisfying the
condition \r{V1} are the absolute velocity (or four-velocity) values.
Their set is denoted by $V(1)$. An inertial world line, i.e.,\ a straight
line, has a constant absolute velocity, a given value $ \vc \in V(1) $. 
Although its elements are vectors, $V(1)$ itself is not a vector space but
only a three dimensional submanifold of $ \vtM / \vtI $. 
 
An observer, usually called also an observer field or a reference frame,
is in reality a collection of pointlike material objects. Any of these
material points embodies a point in the observer's space. Since a material
point is described by a world line, an observer is formulated
mathematically as a continuous collection of world lines, a foliation of
spacetime by a system of world lines. Such a system of world lines can be
given conveniently by the corresponding absolute velocity field, the world
lines being the integral curves of the absolute velocity field. 

This definition allows very general observers, for instance, nonrigid ones
(modelling nonrigid reference media, similarly as, e.g.,\ a fluid or a
dust is used as a reference medium in general relativity), however, here
we will be interested only in inertial observers, for which the velocity
field is a constant $\vc$. Then the world lines of the observer are
parallel timelike straight lines. By means of light signals a
synchronization can be carried out; the synchronous points of the
different world lines prove to form parallel three dimensional spacelike
hyperplanes, these hyperplanes are $\vg$-orthogonal to the world 
lines of the observer.
The elapsed time between two such hyperplanes (actually, two such 
`instants' of the inertial observer) is measured by the proper
time length of the part of any of the world lines falling in between the
two hyperplanes. The system of these hyperplanes gives a foliation of the
spacetime. The world lines of the observer---indeed, the space points of
the observer---prove to form a three dimensional affine space, $\atE_\vc$,
over $\vtE_\vc$, the three dimensional subspace of $\vtM$ which is
$\vg$-orthogonal to $\vc$. The restriction of $\vg$ to $\vtE_\vc$
furnishes $\atE_\vc$ with a Euclidean structure. The hyperplanes (the time
points of the observer) form a one dimensional affine space, $\atI_\vc$,
over $\vtI$. The observer observes a spacetime point $p$ as an $\vc$-time
point (the hyperplane containing $p$) and as an $\vc$-space point (the
world line of the observer that contains $p$). 
 
The inertial observer $\vc$ `splits' a spacetime vector $\vx$ into a time 
interval value and an $\vc$-space vector (an element of $\vtE_\vc$) by 
the projection-like mappings, 
 $$
   \vtau_\vc \, \vx = \vc \cdot \vx, \qquad 
   \vpi_\vc  \vx = \vx - (\vtau_\vc \, \vx) \vc, 
 $$
 respectively. After choosing a spacetime point $o$ as a spacetime origin,
spacetime points also become possible to be described by time intervals
and $\vc$-space vectors: The observer can characterize a spacetime point
$p$ by $ \; \vtau_\vc \, (p-o) \; $ and $ \; \vpi_\vc (p-o) \; $. If, in 
addition, an
orthonormal, positively oriented basis in $ \vtE_\vc / \vtI $ is chosen,
then $\vc$-space vectors can be described by $\vtI$ valued coordinates.
Hence, an origin $o$ and such a basis turns the observer into a
(Cartesian) coordinate system. 
 
The inertial observer $\vc$ observes an arbitrary world line $r$ as a
curve $ r_\vc : \atI_\vc \to \atE_\vc $ in its space. The connection
between an absolute velocity value $\vu \in V(1)$ of the world line $r$
and the corresponding relative velocity value $ v \in \vtE_\vc / \vtI $,
obtained by differentiating $r_\vc$ with respect to its
$\atI_\vc$-variable, is
 $$
   v = \frac{\vu}{ \vc \cdot \vu } - \vc. 
 $$
 
The set of spacetime covectors, $\vtM^*$ (the dual of $\vtM$), can be
identified with $ \vtM / \vtI^\2 $, through the nondegenerate form $\vg$.
Thus $\vtM^*$ inherits the light cone structure that $\vtM$ owns. (It is
this identification through $\vg$ that leads to the possibility to raise
and lower the indices in formulas with indices.)
 
Compared to the special relativistic spacetime, nonrelativistic spacetime
is also considered a (four dimensional, real, oriented) affine space 
$\atM$ (over
the linear space of spacetime vectors, $\vtM$), but in this case an
absolute time / absolute simultaneity structure is assumed. (Actually, it
is not an absolute {\sl space} but an absolute {\sl simultaneity} that is 
needed to
formulate, e.g.,\ gravitational force, as an action-at-a-distance.) Thus,
in addition to $\atM$, a one dimensional oriented real affine space $\atI$
is considered as the set of absolute instants. Its underlying vector
space is $\vtI$, the set of time interval values. Absolute simultaneity is
given by an affine surjection, $ \tau : \atM \to \atI $, which provides a
distinguished system of hyperplanes on $\atM$. One such hyperplane is
formed by those spacetime points $p$ which share the same time value $ t =
\tau (p) $. The linear surjection $ \vtau : \vtM \to \vtI $ that underlies
the affine surjection $\tau$ assigns to any spacetime vector a time
interval, `its time length'. The kernel of $\vtau$, $\vtE \subset \vtM$,
is a three dimensional real oriented vector space. $\vtE$ is the set of
spacelike vectors (the spacetime vectors having a zero time interval).
Nonrelativistically only spacelike vectors have an inner product
structure, a Euclidean one indeed: A positive definite symmetric
bilinear map $ \vb : \vtE \x \vtE \to \vtD^{(2)} $. (Here $\vtD^{(2)}$
expresses the fact that the length of a spacelike vector $\vq$ , $ | \vq |
= \left[ \vb (\vq,\vq) \right]\gyok \: $, has a dimension of length. 
Similarly, 
relativistically $\vg$ maps to $\vtI^\2$, instead of $\vtR$.) Spacetime
vectors which are not spacelike are called timelike, the ones that $\vtau$
assigns a positive or negative time interval value to are called
future-directed or past-directed, respectively. 
 
World lines are again curves in $\atM$ having future-directed tangent
vectors only.  Now it is the absolute time that provides a physically
distinguished, natural parametrization for the world lines. The tangent
vectors obtained by differentiating with respect to this parametrization
prove to be elements of the set $ V(1) := \left\{ \vu \in \vtM / \vtI \; |
\; \vtau \, \vu = 1 \right\} $. The elements of $V(1)$ are the absolute
velocity vectors. Now $V(1)$ is a three dimensional affine subspace of $
\vtM / \vtI $, over $ \vtE / \vtI $. Inertial world lines are again the
ones having a constant absolute velocity, $ \vc \in V(1) $. 
 
Observers, and, in particular, inertial ones, are defined in a way similar
to that in the special relativistic case. Now synchronization is carried
out not by means of light signals but with the aid of absolute time
(because now it is thought that physically an absolute time is available,
`all watches go the same'). All inertial observers use $\tau$ to assign
time points to the spacetime points, thus they do not have different
$\atI_\vc$s but the same $\atI$. On the other hand, different observers 
have
different world line systems, world lines of one observer are not parallel
to the world lines of the other. As a result, the spaces of the observers
will be different; these $\atE_\vc$s prove to be three dimensional
Euclidean affine spaces over the set of spacelike vectors, $\vtE$. 
 
Now an inertial observer observes a spacetime vector $\vx$ as a time
interval $\vtau \, \vx$ and as a $\vc$-space vector $ \: \vpi_\vc \vx = 
\vx - (\vtau \, \vx) \vc. $
Like in the special relativistic case, with a
choice of an origin and a basis, the observer establishes a 
coordinate system; here the basis is chosen in $ \vtE / \vtD $ and
now time coordinates take their values in $\vtI$, while space
coordinates take their values in $\vtD$. 
 
The observer $\vc$ observes an absolute velocity $\vu$ as the relative 
velocity 
   \be{nemseb} 
   \vv = \vu - \vc, \qquad \vv \in \vtE / \vtI,  
   \qquad |\vv| \in \vtD / \vtI . 
   \ee 
In a coordinate system, the time coordinate of an absolute velocity is 
always $1$, and the time coordinate of a relative velocity is always $0$.  
 
Now the set of spacetime covectors (the dual of the linear space of
spacetime vectors), $\vtM^*$, has a structure different from that of 
$\vtM$, 
because now there is no possibility for identification. In $\vtM$ a three
dimensional subspace, $\vtE$, exists as a distinguished subset while in
$\vtM^*$ a one dimensional subspace will be the only distinguished subset.
It is formed by those covectors $\vk$ for which $ \vk \cdot \vq = 0 $ $ (
\minden \vq \in \vtE ) $. These covectors are called timelike and the
others are called spacelike. (Because timelike covectors are the ones that
`examine' the timelikeness of spacetime vectors, i.e.,\ they give zero for
spacelike ones and nonzero for timelike ones.)
 
The Euclidean form $\vb$ provides here an identification not to connect
$\vtM^*$ with $\vtM$ but to connect $\vtE^*$ with $\vtE$. More closely,
the identification is possible between $\vtE^*$ and $ \vtE / \vtD^\2 $. It
is important to observe that $\vtE^*$ is not a subset of $\vtM^*$, but the
restriction of a $ \, \vk \in \vtM^* $, $ \vk |\_\ivtE \: $, is an element of
$\vtE^*$. 

The linear map $ \: \vj : \vtM^* \to \vtE^* $, $ \vk \mapsto \vk |\_\ivtE 
\: $
plays an important role in the $\vc$-splitting of spacetime covectors: an
inertial observer $\vc$ splits a covector $\vk$ into $ \vk \cdot \vc \in
\vtI^* $ and into $ \vj \cdot \vk \in \vtE^* $. If a basis in $ \vtE /
\vtD $ is chosen (to describe spacelike vectors by $\vtD$ valued
coordinates) then, via the dual basis, $ \vj \cdot \vk $ will be possible
to be characterized by coordinates, too (namely, by $\vtD^*$ valued ones). 
 
How tensors and cotensors are split for an observer can also be given with
the aid of $\vtau$, $\vpi_\vc$, and $\vj$. We will make use of the
transformation rules connecting coordinates of vectors, covectors and
cotensors in two different inertial coordinate systems.  If the
same $ \vtE / \vtD $-bases are chosen for the two observers, which 
means physically that the coordinate axes of the two
reference frames are parallel, then the transformation formulas are
   \be{vtr} 
   (x^0)' = x^0, \qquad (x^\alpha)' = x^\alpha - v^\alpha x^0 
   \ee 
for vectors, 
   \be{kvtr} 
   (k_0)' = k_0 + v^\alpha k_\alpha, \qquad (k_\alpha)' = k_\alpha
   \ee 
for covectors, and 
   \[ 
   (C_{00})' = C_{00} + v^\alpha ( C_{0\alpha} + C_{\alpha0} ) +  
   (1/2) v^\alpha v^\beta ( C_{\alpha\beta} + C_{\beta\alpha} ),  
   \] 
   \be{kttr} 
   (C_{\alpha0})' = C_{\alpha0} + v^\beta C_{\alpha\beta}, \qquad 
   (C_{0\beta})' = C_{0\beta} + v^\alpha C_{\alpha\beta}, \qquad 
   (C_{\alpha\beta})' = C_{\alpha\beta} 
   \ee
 for cotensors, where $\vv$ is the relative velocity of the two observers
(our notations are $0$ for timelike, greek letters for spacelike and latin
ones for spacetime indices). For cotensors of higher rank the
transformation formula proves to be a straightforward generalization of
\r{kttr}.


   \section{\point Absolute quantities and equations}


To find absolute quantities and equations equivalent to the wave
function and the Schr\"odinger equation, our starting point is the
Schr\"odinger equation given in the absolute spacetime formalism (cf.\
\cite{MT2}). With the aid of an arbitrary four-velocity value $\vc$ and a
four-potential $\vA$---corresponding to the electromagnetic field strength
$\vF$---, it reads
   \be{Sch1} 
   i \hbar (\vc \cdot D_\vA) \Psi = - 
   \ftort{\fhbarnegyzet}{\fketm}
   (\vj \cdot D_\vA) \cdot (\vj \cdot D_\vA) \Psi 
   \ee
 with $ D_\vA = D\_\atM - i (e/\hbar) \vA $, where $D_\atM$ is the derivation
on the spacetime $\atM$ and $\vA$ denotes the operator of multiplication by
$\vA$ as well. The operators $(e/\hbar) \vA$, $D_\atM$ and $D_\vA$ are
$\vtM^*$ valued vector operators (coordinate-freely, a $\vtV$ valued vector
operator on a Hilbert space $\vtH$ is a linear map $ \vtH \to \vtV \tenzor
\vtH $, see \cite{MT2}). 

The solution space of \r{Sch1} is not only linear
but proves to be a separable Hilbert space (let us denote it by $\vtHH$)
with respect to the scalar product
   \be{skszor}
   \langle \Psi_1, \Psi_2 \rangle = \int_{\atE_t} \Psi_1^* \Psi\_2.
   \ee
 Here integration is performed on one of the spacelike
hyperplanes; from \r{Sch1} it follows that the definition of the 
scalar product does not depend on which hyperplane was chosen. 
 
Four-velocity values and inertial observers are in a one-to-one
correspondence with each other, thus a choice of $\vc$ is equivalent to a
choice of the corresponding inertial observer. That a choice of a $\vc$ is
unavoidable for the Schr\"odinger equation is reflected in that if another
value $\vc'$ is chosen, a solution of \r{Sch1} does not remain invariant.
Instead, it transforms as
   \be{boost1} 
   \Psi_{\vc'}(p)=\Psi_\vc(p) \cdot \exp \left[ i \frac{m}{\hbar} 
   \left(  \frac{\vv^2}{2}\vt - \vv \cdot \vq \right) \right] 
   \ee
 with $\vv=\vc'-\vc$, and with $\vt=\vtau \, (p-p_0)$ as the timelike and
$\vq=\vpi_\vc(p-p_0)$ as the $\vc$-spacelike component of the spacetime
vector $p-p_0$, where $p_0$ is an arbitrary auxiliary spacetime point
(different choices of it mean only constant phase factors, which is
irrelevant). Similarly, a change in the choice of the four-potential, $
\vA' = \vA + D\_\atM \alpha $, involves the transformation of a solution of
\r{Sch1},
 $$
   \Psi_{\vA'} (p) = \Psi_\vA (p) \cdot \exp \left[ i 
   \hspace{.3ex} 
   \ftort{\fe}{\fhbar}
   \hspace{.3ex}
   \alpha (p) \right]. 
 $$
 As a consequence, operators are also observer and gauge dependent. 
Different choices of an observer and a four-potential lead to the
transformation $ O' = T O T^{-1}$ of the operators, where $T$ is the 
following multiplication operator:
 $$
   T = \exp \left[ i \frac{m}{\hbar} \left( \frac{\vv^2}{2} \vt - 
   \vv \cdot \vq + \frac{e}{m} \alpha \right) \right].
 $$
 
After choosing a coordinate system for the inertial observer $\vc$,  
\r{Sch1} turns into  
   \be{Sch2} 
   i \hbar D_0 \Psi = - 
   \ftort{\fhbarnegyzet}{\fketm}
   D_\alpha D_\alpha \Psi, 
   \ee 
 where $ D_k = \parc_k - i (e/\hbar) A_k$. We will use the coordinate form
of Eq.\ \r{boost1}, too. To obtain this, coordinate systems are to be
chosen for both the observers $\vc$ and $\vc'$. If the two observers `use'
the same origin $o \in \atM$ and the same basis in $ \vtE / \vtD $ to
turn themselves into coordinate systems, one finds that the connection
between $\Psi_\vc (\vt, \vr)$ and $\Psi_{\vc'} (\vt', \vr')$ is
   \be{boost2} 
   \Psi_{\vc'} (\vt', \vr') = \Psi_\vc (\vt', \vr' + \vv \vt') \cdot  
   \exp \left[ i \frac{m}{\hbar} \left( - \frac{\vv^2}{2} \vt' - 
   \vv \cdot \vr' \right) \right]. 
   \ee 
 
We want to find absolute quantities instead of the relative quantity
$\Psi$. To this end, let us consider the polar decomposition of the wave
function, $\Psi=Re^{iS}$, where $R$ and $S$ are real functions. $R$ is an
absolute quantity as boosts, gauge transformations and the ray
indefiniteness all touch only the phase of the wave function. From $S$ we
constitute
   \be{edef} 
   \ep := 
   \ftort{\fhbar}{\fm}
   \parc_0 S - 
   \ftort{\fe}{\fm}
   A_0  
   \ee 
and  
   \be{udef} 
   u_\alpha := 
   \ftort{\fhbar}{\fm}
   \parc_\alpha S - 
   \ftort{\fe}{\fm}
   A_\alpha. 
   \ee  
 The quantities $\ep$ and $u_\alpha$ are also gauge independent as well
as ray ambiguity free. The definitions \r{edef}, \r{udef} imply the
following consistency conditions: 
   \begin{eqnarray}\l{Efelt} 
   \parc_\alpha \ep - \parc_0 u_\alpha = & \! \! \! -
   \ftort{\fe}{\fm}
   F_{\alpha 0} \! \! \! & = - 
   \ftort{\fe}{\fm}
   E_\alpha, 
   \\ \l{Bfelt}  
   \parc_\alpha u_\beta - \parc_\beta u_\alpha = & \! \! \! -
   \ftort{\fe}{\fm}
   F_{\alpha\beta} \! \! \! & = - 
   \ftort{\fe}{\fm}
   \epsilon_{\alpha\beta\gamma} B_\gamma,
   \end{eqnarray} 
 where the observer $\vc$ splits the electromagnetic field tensor $\vF = d
\vA$ into the electric field $E$ and the magnetic field $B$. Conversely,
the equations \r{Efelt}, \r{Bfelt} are sufficient conditions for the way
in which $\Psi$ can be reconstructed from $R$, $\ep$ and $u_\alpha$,
   \be{Psiintegral} 
   \Psi (p) = R \cdot \exp \left[ i \left\{ 
   \ftort{\fm}{\fhbar}
   \int_o^p \left[ \left( \ep + 
   \ftort{\fe}{\fm}
   A_0 \right) dx^0 + \left( u_\alpha + 
   \ftort{\fe}{\fm}
   A_\alpha \right) dx^\alpha \right] + S(o) \right\} \right],  
   \ee 
 to be independent of the path of integration. The constant $S(o)$ in 
\r{Psiintegral} remains undetermined, which means indefiniteness up to a 
constant phase factor. This is just in accord with the fact that not wave 
functions but 
rays bear a physical meaning. Thus we see that the quantities $R$, $\ep$ 
and $u_\alpha$ prove to be equivalent to the wave function---from the 
physical point of view, i.e.,\ they are equivalent to the ray whose 
representative is $\Psi$.
 
The quantities $\ep$ and $u_\alpha$ are derived with the aid of a
coordinate system. To see what coordinate and observer free quantities can
be found behind them, let us compute their covariance properties under a
change of the observer. A quick look at \r{edef} and \r{udef} might
suggest that $\ep$ and $u_\alpha$ form a covector, but actually they do
not, since $S$ is not boost invariant [cf.\ \r{boost1} or \r{boost2}].
Instead, we find that they are transformed as
 $$
   \ep' = \ep + v^\alpha u_\alpha - (1/2) \vv^2, \qquad \qquad
   u_\alpha' = u_\alpha - v^\alpha. 
 $$
 Hence, first, $u_\alpha$ can be regarded as the space components of an
absolute velocity field $u$: 
 $$
   u^0 := 1, \qquad u^\alpha := u_\alpha
 $$
 [cf.\ \r{vtr}, and the remark made after \r{nemseb}]. It is easy to see
that the probability four-current, which, in terms of the wave function, 
reads
 $$
   j^0 := \Psi^* \Psi,   \qquad   j^\alpha := 
   \ftort{\fhbar}{\fketim}
   \left[ \Psi^* \parc_\alpha \Psi - \Psi \parc_\alpha \Psi^* \right],
 $$
 can be expressed 
with this $u$ as $ j = R^2 u $.  Next, from $\ep$ the scalar quantity
   \be{sdef}
   s := - \ep - (1/2) u_\alpha u_\alpha
   \ee
 can be formed. With these quantities the Schr\"odinger equation, imposed
on the complex wave function, proves to be equivalent to the following two
real equations: 
   \ba{lap}
    & s R + \ftort{\fhbarnegyzet}{\fketmnegyzet} \Laplace R = 0,
    & \\ \l{kont}
   & \mbox{Div} j = 0. &
   \ea
 Concerning the first equation we remark that nonrelativistically the
Laplacian, in a coordinate system, $ \Laplace = \parc_\alpha \parc_\alpha
$, is an absolute operation. Expressing \r{Efelt} and \r{Bfelt} with $u$
and $s$, in the absolute form, yields
   \be{Efelt2}
   u \cdot D\_\atM u + \vj \cdot D\_\atM s = 
   \ftort{\fe}{\fm}
   \: \vj \cdot F u
   \ee
and
   \be{Bfelt2}
   \left( \vj D_\atM u \right)^T - \left( \vj D_\atM u \right) = 
   \left( \vj \tenzor \vj \right) \cdot F.
   \ee
 In a coordinate system \r{Efelt2} reads
   \be{Efelt3}
   \parc_0 u_\alpha + u_\beta \parc_\beta u_\alpha + \parc_\alpha s = 
   \ftort{\fe}{\fm}
   \left( E_\alpha + \epsilon_{\alpha \beta \gamma} u_\beta B_\gamma 
   \right).
   \ee
 Then an absolute process can be given by the quantities $R$, $u$, and
$s$, satisfying the absolute equations \r{lap}, \r{kont}, \r{Efelt2}, and
\r{Bfelt2}. Naturally, wherever $R$ will be mentioned, $ \ro = R^2$ can be
taken, instead, as well. The quantity $\ro$ has a direct physical
meaning, as expressing the position probability density. 

We remark that in the hydrodynamical formalisms \cite{Mad,Tak,Jan,BH,KR},
usually the quantities $R$, $u_\alpha$, and $s$, understood with the
corresponding equations, are used. $s$ is usually called the ``quantum
potential'', because of its appearance in \r{Efelt3} (usually it is placed 
to the r.h.s.\ as $ - \parc_\alpha s $). We do not use this misleading name
because $s$ is not an outer, given quantity but a variable, it is a part
of the absolute process describing the particle. 

Alternatively, $\ep$ and $u_\alpha$ can be viewed as components of a 
two-cotensor: 
   \be{z}
   z_{00} = \ep, \qquad z_{\alpha 0} = z_{0 \alpha} = \fel u_\alpha, 
   \qquad z_{\alpha \beta} = - \fel \delta_{\alpha \beta}
   \ee
[cf.\ \r{kttr}], or components of a three-cotensor:
   \ba{w}
   & w_{000} = \ep, \qquad w_{\alpha 00} = u_\alpha, \qquad 
   w_{0 \alpha 0} = w_{00 \alpha} = 0, & \nn
   & w_{\alpha \beta 0} = w_{\alpha 0 \beta} = - w_{0 \alpha \beta} =
   - \frac{1}{2} \delta_{\alpha \beta}, \qquad w_{\alpha \beta \gamma} 
   = 0. &
   \ea
 This latter cotensor provides an elegant way to formulate the equations 
\r{Efelt} and \r{Bfelt} as
   \be{wegy2}
   \parc\_k w\_{l00} - \parc\_l w\_{k00} = - 
   \ftort{\fe}{\fm}
   \, F_{kl}.
   \ee
 The coordinate free form of this equation is
   \be{wegy}
   D\_\atM \ek w = - 
   \ftort{\fe}{\fm}
   \: F \tenzor ( \vtau \tenzor \vtau ),
   \ee
 where in $ D\_\atM \ek w = \left( D\_\atM w \right)^T - D\_\atM w \: $ 
transposition
occurs in the first and last variables of the four-cotensor. We remark
that $ (\vtau \tenzor \vtau)_{mn} = \delta_{m0} \delta_{n0} $ in any
inertial coordinate system. 

The quantity $s$ can be considered as an invariant scalar of the cotensor
$w$. Similarly, $u$ is an invariant four-vector of $w$. Furthermore, from
$ W := \ro w $ all the quantities found above ($\ro$, $u$, $s$, $j$, $z$,
$w$, and, in a reference frame, $\ep$) can be recovered. Therefore, in the
following we will refer to an absolute process as $W$, a process
function which is a three-cotensor field. 

The complete set of absolute equations is formed by \r{lap}, \r{kont} and
\r{wegy}. Eq.\ \r{wegy} is equivalent to the two equations \r{Efelt2}
and \r{Bfelt2}, thus \r{lap}, \r{kont}, \r{Efelt2} and \r{Bfelt2} also
give a complete set of equations. The set of process functions $W$
satisfying the absolute equations is the absolute process space
(denoted by $\cS$). 

Let us investigate the physical meaning of the absolute equations. 
Eq.\ \r{kont} is the continuity equation for the probability
four-current. Eq.\ \r{wegy} tells how the outer field excerts its
action on the particle. In Sect.~X. we will see that relativistically a
similar equation will hold for a covector quantity. The
nonrelativistic limit of the components of that covector will lead to the
components of the nonrelativistic $w$. At last, \r{lap} is the
nonrelativistic and quantum appearance of the mass shell condition for 
the particle. 
To see this, let us recall that in relativistic classical mechanics the
connection between kinetic energy and momentum of a point particle of mass
$m$ is the mass shell condition $ p^k p_k = m^2 $. In Sect.~X. the
corresponding quantum equation will be presented. In nonrelativistic
classical mechanics this mass shell condition is $ 2 m K - p_\alpha
p_\alpha = 0 $ $\;$($ K = p_\alpha p_\alpha / 2m $). As we will see in
Sect.~VI., $\ep$ and $u_\alpha$ are related to the kinetic energy, 
respectively the velocity, physical quantities of the particle. This, 
together with
\r{sdef} and the corresponding relativistic formula \r{rel2}, makes it
transparent that Eq.\ \r{lap} can be interpreted as the quantum version 
of the nonrelativistic mass shell condition. 

We remark that, in parallel to relativistic classical mechanics, where the
Newton equation gives the (proper) time derivative of the kinetic
energy-momentum four-vector, in nonrelativistic classical mechanics, 
kinetic energy and momentum form a cotensor
[similarly as $\ep$ and $u$ form $w$, cf.\ \r{w}], and the equation which
gives how the time derivative of this cotensor is determined by the outer
force is equivalent to the Newton equation. 

We can observe that, instead of a single equation, the Schr\"odinger 
equation,
absolutely a system of equations are present. Actually, this happens to be
an advantage of the formalism. Namely, it expresses that different
elements of the system exist, which are independent. The continuity
equation, the way the outer field acts on the particle, and the mass shell
condition are physically independent aspects of the system. As an example,
in Sect.~VII. we will see how the absolute formalism can be extended to be
valid for not only electromagnetic (or other conservative) force fields by
generalizing \r{wegy}, while keeping the two other equations untouched. In
contrast, in the Schr\"odinger equation these three elements appear in
a mixed and more hidden form.


   \section{\point The structure of the absolute process space}


Let us now examine the geometric properties of the absolute process space,
$\cS$. To do this, the connection with the Hilbert space $\vtHH$ will be
useful. However, most elements of the structure of $\vtHH$, including the
scalar product, the distance function defined as the norm of the
difference of two elements, and the topology defined by the distance
function, are not absolute: They are not ray ambiguity free. To explore
the absolute process space, only elements with absolute meaning can be
made use of. 

The most important of them is the magnitude of the scalar product. This 
quantity possesses direct physical meaning. We introduce the notation
   $$ 
   \sa ( \vW_1, \vW_2 ) := | \langle \Psi_1, \Psi_2 \rangle |, 
   $$
 where $\vW_i$ is the absolute process corresponding to the normalized
wave function $\Psi_i$. {}From \r{skszor} and \r{Psiintegral}, the
explicit form of $ \sa ( \vW_1, \vW_2 ) $ in terms of absolute quantities
can be obtained easily (here we will not need the concrete formula). 

First, let us see how, with the aid of the function $\sa$, $\cS$ becomes a
topological space. For any element, $\vW_0$, of $\cS$, and any number $
\xi \in [0, 1) $, let us define the neighbourhood $B_\xi (\vW_0)$ as
formed by those processes $\vW$ for which $ \sa ( \vW_0, \vW ) > \xi $.
Then, by taking the topology generated by all $B_\xi(\vW_0)$s, $\cS$
is made a topological space. 

As the next step, we show that the absolute process space is a metrizable 
space. To see this, let us look for all such possible distance functions
on $\cS$ that are compatible with the given topology, in other words, that
define the same neighbourhoods (their open balls) as were defined before
through $\sa$. $\sa$ itself is not such a distance function, and $\xi$ 
is not a radius-like quantity of $B_\xi(\vW_0)$, because $ \sa (\vW, 
\vW) = 1 $ instead of being zero, and `bigger and bigger' $B$s have smaller 
and 
smaller $\xi$s (if $ B_{\xi_1}(\vW_0) \subset B_{\xi_2}(\vW_0) $ then $ 
\xi_1 > \xi_2 $). Based on this observation and the expected properties 
of a distance function, we look for all such strictly monotonously 
decreasing non-negative functions $f$ for which the distance function
$ \dl := f \kor \sa $ satisfies the triangle inequality.

One simple way to find all such possible $f$s is to work in terms of wave
functions, instead of absolute processes. Then the triangle inequality
reads as the condition
   \be{haromszog}
   f ( | \langle \Psi_1, \Psi_2 \rangle | ) + 
   f ( | \langle \Psi_1, \Psi_3 \rangle | ) \geq 
   f ( | \langle \Psi_2, \Psi_3 \rangle | )
   \ee 
 for any three wave functions of unit norm. By carrying out a
Schmidt-orthogonaliz\-ation, the relative positions of three wave functions
can be characterized by some angle parameters.  By means of such
parameters, the absolute values of the three scalar products in question
can be expressed as 
   $$
   | \langle \Psi_1, \Psi_2 \rangle | = \cos \alpha, \qquad
   | \langle \Psi_1, \Psi_3 \rangle | = \cos \beta, 
   $$
   $$ 
   | \langle \Psi_2, \Psi_3 \rangle | = | \cos \alpha \cos \beta + 
   \sin \alpha \sin \beta \cos \gamma e^{i\delta} |,
   $$
 where $ \alpha, \beta, \gamma \in [0, \pi/2] $ and $ \delta \in [0, 2\pi)
$. After maximizing the r.h.s. of \r{haromszog}, first in $\delta$, and
then in $\gamma$, one finds that, with fixed $\alpha$ and $\beta$, the
strongest condition \r{haromszog} imposes is $ f ( \cos \alpha ) + f (
\cos \beta ) \geq f [ \cos ( \alpha + \beta ) ] $.  It is useful to
introduce $ h := f \kor \cos $, with which the condition reads
   $$
   h(\alpha) + h(\beta) \geq h(\alpha + \beta). 
   $$

Now let us recall that, on any metric space with a distance function
$\dl$, $g \kor \dl$ is also a distance function and defines the same open
balls, if $g$ is such a strictly monotonously increasing function that
$g(0)=0$ and $ g(a+b) \leq g(a) + g(b) \; (\minden a,b \in \vtR^+_0) $.
This degree of nonuniqueness of the distance function is natural on
any metrizable topological spaces. 

We see that if we choose $ h(\alpha) = \alpha $ (which is the
most natural choice of $h$), i.e.,\ we choose the distance function $ \dl
= \arccos \sa $, then the other allowed distance functions differ from it
just to the extent of such a $g$ arbitrariness. Therefore, to the extent
any distance function can be unique on a metrizable topological space, $ 
\dl = \arccos \sa
$ is the unique distance function on $\cS$ that satisfies our
requirements. In what follows, we will consider $\cS$ a metric space with
respect to the distance function $ \dl = \arccos \sa $. 

 Concerning the question of completeness of the metric space $\cS$, we
remark that the absolute equations, just like the Schr\"odinder equation,
are valid only for sufficiently smooth process functions. One can prove 
that the
absolute version of how one makes $\vtHH$ complete by means of Cauchy
sequences of smooth wave functions is just how one makes $\cS$ complete by
means of Cauchy sequences, here with respect to the absolute distance
function $\dl$. 

Our next observation is that the absolute process space is a (complex)
Hilbert manifold. This can be seen with the aid of the following
homeomorphism between a neighbourhood $B_\xi(\vW_0)$ and a neighbourhood
of a Hilbert space: After fixing an $\vc$ and $\vA$, a ray $\ray_0$ of
$\vtHH$ corresponds to $\vW_0$. Let us choose an element $\Psi_0$ from
this ray. Similarly, let us choose a $\Psi$ from the ray $\ray$
corresponding to an arbitrary $\vW$ belonging to the neighbourhood
$B_\xi(\vW_0)$ of $\vW_0$. Then
   \be{fidef} 
   \varphi := \frac{ | \langle \Psi_0, \Psi \rangle | }{ 
   \langle \Psi_0, \Psi \rangle }
   [ \Psi - \langle \Psi_0, \Psi \rangle \Psi_0 ] 
   \ee
 is such a vector that, on the one hand, is orthogonal to $\Psi_0$, and,
on the other hand, does not depend on how $\Psi$ was chosen from the ray
$\ray$. This means that, actually, $\varphi$ characterizes $\ray$, and,
correspondingly, $\vW$. The map \r{fidef} proves to provide a
homeomorphism between $B_\xi(\vW_0)$ and $ \{ \varphi \in \vtHH \: | \;
\varphi \ortog \Psi_0, \; \| \varphi \| < (1 - \xi^2)\gyok \} $, the
latter being a neighbourhood of a separable complex Hilbert space. 

Based on the parametrization \r{fidef}, the tangent vectors of $\cS$ at
$\vW$ can be brought to a linear one-to-one correspondence with those
vectors of $\vtHH$ which are orthogonal to $\Psi_0$. Moreover, if, using
this correspondence, we `transport' the scalar product of $\vtHH$ to the
absolute tangent space, we obtain an absolute scalar product on it. As one
can check, this scalar product remains invariant under different choices
of $\Psi_0 \in \ray_0$, or different choices of $\vc$ or $\vA$. This way
the absolute process space proves to be a complex Hilbert manifold. 

Similarly to the treatment that can be found in \cite{AS} (where 
the manifold of the rays of a Hilbert space is studied), this complex
Hilbertian structure of $\cS$ can be replaced by a K\"ahler structure. 
This alternative aspect is the following: Let us consider a tangent space
of $\cS$ a real vector space, instead of a complex one. The multiplication
with $i$ is replaced by the corresponding real linear operator $J$ (having
the property that $ J^2 = - I $, where $I$ is the identity operator).
Then, from the complex scalar product $ \langle \varphi, \psi \rangle $,
we obtain a (real) Riemannian form, 
   $$
   G ( \varphi, \psi ) := \mbox{\sl Re} \langle \varphi , \psi \rangle, 
   $$
 and a (real) symplectic form, 
   $$
   \Omega ( \varphi, \psi ) := 
   - \mbox{\sl Im} \langle \varphi , \psi \rangle. 
   $$ 
 These two forms are related to each other as
   \be{viszony}
   G ( \varphi, \psi ) = \Omega ( \varphi, J \psi ). 
   \ee
 Hence, this way $\cS$ can be viewed as a K\"ahler manifold: a real,
Riemannian and symplectic manifold, having a complex structure and the
property \r{viszony}. 

This alternative aspect can be useful, for example, if one wants to
investigate the connection between the absolute quantum process space and
the corresponding classical one. As we have mentioned, the classical
process space is a symplectic manifold, thus the symplectic structure 
of the quantum process space may provide a
transparent connection between the classical system and the quantum one.
In terms of phase spaces such an investigation has already been carried
out by \cite{AS}. Hopefully, a similar connection will turn out for the 
process spaces, the absolute equivalents of phase spaces, too.

Finally, we show that the distance function that can be defined by $G$
coincides with the distance function $\dl$ found above. A way to see this
is to determine the $G$-shortest curve between two points with calculus of
variation, and to compare its $G$-length with the $\dl$-distance of the
two points. With the above notations, the $G$-length of a curve leading
from $\vW_0$ to $\vW$, in the coordinatization \r{fidef} (obtained after a
choice of an $\vc$, an $\vA$, and a $\Psi_0$), turns out to be
   \be{hossz}
   \int_0^1 \left\{ \| \dot{\varphi} \|^2 + \frac{1}{ 1 - 
   \| \varphi \|^2 } \left[ \mbox{\sl Re} \langle \varphi(t), 
   \dot{\varphi}(t) \rangle \right]^2 - 
   \left[ \mbox{\sl Im} \langle \varphi(t), \dot{\varphi}(t) \rangle
   \right]^2 \right\}\gyok \; dt
   \ee
 ($ \varphi(0) = 0 $ is the coordinate of $\vW_0$, and $\varphi(1)$ is the
coordinate of $\vW$). By deriving and solving the corresponding
Euler-Lagrange equation, after lengthy but not particularly illuminating 
calculations,
the geodesic between the two points can be given in the following very
simple form: 
   \be{geod}
   \varphi(t) = t \cdot \varphi(1), \qquad t \in [0,1]
   \ee
 [after fixing the reparametrization ambiguity stemming from the
reparametrization invariance of the integral \r{hossz}]. The $G$-length
of this geodesic can then be calculated easily, and the result is $ \: 
\arccos | \langle \Psi_0, \Psi \rangle |, \,$ which is exactly $ \dl (
\vW_0, \vW ) $. 

As we see, the form of the geodesic \r{geod} is just the one one 
expects on intuitive grounds. This also shows an advantage of the 
K\"ahlerian aspect of the geometry of the process space.


   \section{\point Event space and states}


The investigation of the mathematical description of the events of quantum
mechanics and the relation of the event space to the states and physical
quantities of the system was started by Birkhoff and Neumann. These
results and later developments (see \cite{Mac,Seg,Gud,MT2}) led to the
following picture: There is a Hilbert space $\vtH$ for the system in such
a way that events are projections of $\vtH$, states are density operators
on $\vtH$, and physical quantities are self-adjoint operators on $\vtH$.
Or, equivalently, as density operators are equivalent to probability
distributions (probability laws) on the space of events, self-adjoint
operators (through their spectral resolution) are equivalent to projection
valued measures, and projections are in a one-to-one correspondence with
the (closed) linear subspaces of $\vtH$, the picture can be expressed
in the following way: Events are the linear subspaces,
states are probability distributions on the event space, and physical
quantities are event valued measures. 

This latter scheme is a close parallel of the corresponding situation in
classical mechanics (\cite{Mac,MT2}). There events are subsets of the
phase space, or, in the absolute treatment, of the process space, states
are probability distributions on the event space, and physical quantities
are event valued measures (as equivalents of functions defined on the
phase space or the process space). 

The structure of the classical mechanical event space is as follows.
The operation `and' between two events (subsets) $A$ and $B$ is $ A \es B
:= A \metszet B $, the operation `or' is $ A \vagy B := A \unio B $, the
`negation' $A^\ortog$ is the set theoretical complement, and the relation $
A \leq B $ is the relation $ A \subseteq B $. (Here we do not wish to
go into mathematical details, see \cite{MT2} concerning them.) In quantum
mechanics, `and' is the intersection of linear subspaces, `or' of $A$
and $B$ is the linear subspace spanned by $A$ and $B$, the `negation' is 
the
orthocomplement ($A^\ortog$ consists of all the elements of $H$ being
orthogonal to $A$), and the relation $ A \leq B $ is the relation $ A
\subseteq B $. (Again, see \cite{MT2} for details.)

While in classical mechanics the event space derived from the process
space is absolute, in quantum mechanics, even if working from $\vtHH$ as
the Hilbert space, the resulting events are relative. In the following we
establish the absolute event space (it will be denoted by $\cE$) based on
the absolute process space $\cS$. It will be easy to check that, based on
\r{Psiintegral}, both the events of $\cE$ and the operations on $\cE$ can
be brought into one-to-one correspondence with the linear subspaces of
$\vtHH$ and the above mentioned operations among them, respectively. Thus
the absolute event space expresses completely the same physics as the one
built from $\vtHH$ does. 

The key ingredient in constructing the absolute event space from $\cS$ is
the following relation on $\cS$: Let us call two elements, $W_1$ and
$W_2$, orthogonal to each other if $ \sa (W_1,W_2) = 0 $. In terms of
the distance function $\dl$ this means that those elements of $\cS$ are
orthogonal to each other which have the greatest distance `available' on
$\cS$. Furthermore, for any subset $A$ of $\cS$, let $A^\ortog$ denote the
set of all such elements of $\cS$ that are orthogonal to all elements of
$A$. With the aid of this, we introduce events as such subsets of $\cS$ 
that satisfy the condition $ A = (A^\ortog)^\ortog $. 

Then, the operations on $\cE$ formed by these events are defined as
follows. Let the `and' of $A$ and $B$ be $ A \es B := A \metszet B $, let
the `negation' of $A$ be $A^\ortog$, let the `or' be defined through the 
de Morgan formula: 
 $$
   A \vagy B := ( A^\ortog \es B^\ortog )^\ortog,
 $$
 and let the relation $ A \leq B $ be the relation $ A \subseteq B $.

As mentioned, it is easy to see that this event space with these
operations is actually just the absolute counterpart of the one built from
$\vtHH$. This, at the same time, ensures that the introduced operations
satisfy the necessary properties so we do not need to check them directly. 

In the absolute form of classical mechanics elementary events are those
subsets of the process space which contain only one element, i.e.,\
processes are the elementary events. In the event space of Hilbert space
quantum mechanics elementary events are the one dimensional linear
subspaces, or equivalently, rays, of the Hilbert space. In $\cE$ 
subsets containing only one element are the elementary events. In 
other words, here also the processes are the elementary events.

Now, having established the absolute event space, we can introduce states
as the probability distributions (probability measures) on it. As special
examples, pure states are such states that take the value $1$ on one
elementary event. As is known from the Hilbert space formalism of quantum
mechanics, a pure state is actually uniquely determined by the elementary
event on which it takes the value $1$. Thus a process---an element of the
process space---can be regarded as an elementary event and as a
pure state, too. In the pure state $p_\sW$ belonging to the process $W$,
the probability of an elementary event $A_{\sW'}$ naturally turns out to
be $\sa^2(W,W')$. Mixed states can be given as convex combinations $
\sum_{n \in N} \lambda_n p_n $ of pure states---this property is the
absolute parallel for the corresponding property of density operators
\cite{MT2}.


   \section{\point Physical quantities}


Next, we formulate physical quantities, also with respect to $\cE$. We
illustrate how the general `machinery' for giving physical quantities as
event valued maps (established in \cite{Mac} and developed in \cite{MT2})
works here through the example of the position physical
quantity. 

Position, which is a vector operator in the Hilbert space language, is a
physical quantity that is measured at a given instant with respect to an 
inertial observer, supplied with a space origin. 
If, at a time $t$, the particle is found in a subset $B$ of the
space of the observer $\vc$ (a detector, being at rest with respect to the
observer $\vc$, observes the particle at $t$ in the space volume $B$),
then this observation is an event of the particle. To express the fact
that these position observations are (certain) events of the considered
physical system, we formulate position as a map from the subsets of the
observer's space to the events of the system. 

Namely, this map (let us denote it by $Q_{\vc,t}$) is the following: For a
given instant $t$, let us assign to a subset $B$ the event formed by those
processes $\vW$ for which
   \be{helydef}
   \ro_\sW (t, \br) = 0 \qquad ( \barmely \br \not\in B ). 
   \ee

As in the general framework, a state, i.e.,\ a probability distribution on
the events, implies probability distributions for each physical quantity.
In a state $p$, the probability distribution for a physical quantity $F$
is $ p \kor F $. For example, in the case of the position, $ p \kor
Q_{\vc,t} $ is a probability distribution on the observer's space. This
formula is transparent: To a subset $B$ in the observer's space,
$Q_{\vc,t}$ assigns the event $ Q_{\vc,t} (B) $, and $ p(Q_{\vc,t} (B)) $
tells the probability for this event, and consequently the probability
belonging to $B$. This value is the probability for finding the particle
in $B$ at $t$. 

In the case $p$ is a pure state $p_\sW$, one finds that $ p ( Q_{\vc,t}
(B) ) = \int_B \ro_\sW ( t , \br ) d^3 \br $. As expected, this result
tells nothing else but that, in a pure state belonging to the process $W$,
$\ro_\sW$ expresses the probability density of the position. 

Other physical quantities are formulated similarly. Each
physical quantity is a map from the subsets of a vector space to the
event space, where the vector space in question is the one that the 
measured values 
of the physical quantity are elements of. For example, in the case of
the velocity physical quantity, i.e.,\ the relative velocity with respect
to an observer $\vc$ and an instant $t$, the possible velocity values
are the elements of $ \vtE / \vtI $ [cf.\ \r{nemseb}]. Correspondingly,
the velocity physical quantity (notation: $\VV$) is a map assigning 
events to the subsets of $ \vtE / \vtI $. 

There exists an alternative way to give physical quantities. This other
possibility is based on the property that a physical quantity is uniquely
determined by its expectation values computed in all the pure states. 
This fact is the absolute version of the corresponding statement in a
Hilbert space formalism, where it is not hard to show that any matrix
element of a self-adjoint operator can be determined from some appropriate
expectation values of it. 

In the case of the velocity physical quantity, which in a Hilbert space
formalism leads to the (gauge invariant) velocity component operators $
\frac{\hbar}{im} \parc_\alpha - \frac{e}{m} \vA_\alpha $, the expectation
value in a pure state $p_\sW$ proves to be
   \be{sebvarh}
   \langle \VV \rangle = 
   \int \ro ( u - \vc ) ,
   \ee
or, in coordinates,
   \be{sebkompvarh}
   \langle (\VV)_\alpha \rangle = 
   \int \ro u_\alpha  = \int j_\alpha,
   \ee
 where $\ro$, $u$ and $j$ are the quantities belonging to $W$, and the
integrations are performed in the observer's space at the time point $t$. 
The formula \r{sebvarh} [or \r{sebkompvarh}] provides a convenient way to
give the velocity physical quantity in the absolute formalism. Similarly,
we give the kinetic energy physical quantity $\KK$ also by means of its
expectation values in pure states. We find
   \be{envarh}
   \langle \KK \rangle = - m \int \ro \ep = m \int \ro \left( 
   \ftort{\fegy}{\fketto} \hspace{.2ex}
   u_\alpha u_\alpha + s \right).
   \ee

Now we can see the physical meaning of the quantities $u_\alpha$ and
$\ep$: They express the kinetic energy and the kinetic momentum of the
particle. 

If we calculate the uncertainty (standard deviation) of a velocity
component, the result is
   \be{sebszor}
   \left[ \Delta \left( \VV \right)_\alpha \right]^2 = 
   \int \ro \left[ u_\alpha - \langle \left( \VV \right)_\alpha \rangle 
   \right]^2 + 
   \ftort{\fhbarnegyzet}{\fmnegyzet}
   \int \left( \parc_\alpha R \right)^2
   \ee
 (here and in the following formulas of this section $\alpha$ is a fixed
index, its double occurences do not imply summation). This formula offers
to obtain some sharper inequalities than the usual position-velocity
uncertainty formula. [As mentioned before, we prefer velocity to canonical
momentum in uncertainty inequalities because canonical momentum is a gauge
dependent quantity while velocity is not. From the aspect of the
Heisenberg uncertainty inequality they behave the same way, which can be
seen, for example in the Hilbert space formalism, from that---as
operators---their commutator with position are the same (except for a
factor $m$).]

To derive these sharper inequalities, by using the
Cauchy-Bunyakovski-Schwartz inequality for real functions,
 $$
   \int f^2 \; \cdot \int g^2 \geq \left\{ \int f g \right\}^2,
 $$
we find
   \begin{eqnarray}
   \int \left\{ R \left[ r_\alpha - \langle \left( Q_{\vc,t} 
   \right)_\alpha \rangle \right] \right\}^2 \cdot 
   \ftort{\fhbarnegyzet}{\fmnegyzet}
   \int \left\{ \parc_\alpha R \right\}^2 \geq 
   \qquad \qquad \qquad \qquad \qquad \nn \l{CSBb}
   \qquad \qquad \qquad \qquad
   \ftort{\fhbarnegyzet}{\fmnegyzet}
   \left\{ \int \left[ r_\alpha - \langle 
   \left( Q_{\vc,t} \right)_\alpha \rangle \right] \parc_\alpha \left( 
   \ftort{\fegy}{\fketto}
   R^2 \right) \right\}^2 = 
   \ftort{\fhbarnegyzet}{\fnegymnegyzet}
   ,
   \end{eqnarray}
and
   \begin{eqnarray}
   \int \left\{ R \left[ r_\alpha - \langle \left( Q_{\vc,t}
   \right)_\alpha \rangle \right] \right\}^2 \cdot
   \int \left\{ R \left[ u_\alpha - \langle \left( \VV \right)_\alpha 
   \rangle \right] \right\}^2 \geq
   \qquad \qquad \qquad \nn \l{CSBa}
   \qquad \qquad \qquad
   \left\{ \int R^2 \left[ r_\alpha - \langle \left( Q_{\vc,t}
   \right)_\alpha \rangle \right] \left[ u_\alpha - \langle 
   \left( \VV \right)_\alpha \rangle \right] \right\}^2
   \end{eqnarray}
 (here and in the following surface terms are always dropped). Both the
l.h.s.\ of \r{CSBb} and the l.h.s.\ of \r{CSBa} are less than or equal to $ 
\left[ 
\Delta \left( Q_{\vc,t} \right)_\alpha \right]^2 \left[ \Delta \left( \VV
\right)_\alpha \right]^2 $ [cf.\ \r{sebszor}]. Furthermore, the sum of 
these 
two l.h.s.-s is just $ \left[ \Delta \left( Q_{\vc,t} \right)_\alpha
\right]^2 \left[ \Delta \left( \VV \right)_\alpha \right]^2 $. Thus we
find the following three inequalities: 
   \begin{eqnarray}\l{hat1}
   \left[ \Delta \left( Q_{\vc,t} \right)_\alpha \right]^2 
   \left[ \Delta \left( \VV \right)_\alpha \right]^2 & \geq &
   \ftort{\fhbarnegyzet}{\fnegymnegyzet}
   + \left[ \Delta \left( Q_{\vc,t} \right)_\alpha \right]^2 \cdot
   \int \left\{ R \left[ u_\alpha - \langle \left( \VV \right)_\alpha 
   \rangle \right] \right\}^2, 
   \\ \l{hat2}
   \left[ \Delta \left( Q_{\vc,t} \right)_\alpha \right]^2 
   \left[ \Delta \left( \VV \right)_\alpha \right]^2 & \geq &
   Y_\alpha^2 +
   \left[ \Delta \left( Q_{\vc,t} \right)_\alpha \right]^2 \cdot
   \ftort{\fhbarnegyzet}{\fmnegyzet}
    \int \left\{ \parc_\alpha R \right\}^2, 
   \\ \l{hat3}
   \left[ \Delta \left( Q_{\vc,t} \right)_\alpha \right]^2 
   \left[ \Delta \left( \VV \right)_\alpha \right]^2 & \geq &
   \ftort{\fhbarnegyzet}{\fnegymnegyzet}
   + Y_\alpha^2
   \end{eqnarray}
with
   \be{Y}
   Y_\alpha = \int R^2 \left[ r_\alpha - \langle \left( Q_{\vc,t}
   \right)_\alpha \rangle \right] \left[ u_\alpha - \langle
   \left( \VV \right)_\alpha \rangle \right].
   \ee
 {}From these inequalities \r{hat1} and \r{hat3} are stronger (or, at 
least, not weaker) than the usual uncertainty formula $ \Delta
\left( Q_{\vc,t} \right)_\alpha \Delta \left( \VV \right)_\alpha \geq 
\hbar/2m $, while \r{hat2} is an independent inequality.

Finally we prove Ehrenfest's theorem, in terms of the absolute formalism. 
First calculate the first time-derivative of the expectation value of the
position of the particle: 
   \be{rpont1}
   \frac{\partial}{\partial t} \langle \left( Q_{\vc,t} \right)_\alpha 
   \rangle = \frac{\partial}{\partial t} \int \ro r_\alpha = 
   \int {r_\alpha \frac{\partial \ro}{\partial t} }.
   \ee
Using the continuity equation, \r{kont}, we get
   \be{rpont2}
   \frac{\partial}{\partial t} \langle \left( Q_{\vc,t} \right)_\alpha 
   \rangle = \langle \left( \VV \right)_\alpha \rangle.
   \ee
The second derivative of the expectation value of the position can be
written now as
   \be{rpontpont1}
   \frac{\partial^2}{\partial t^2} \langle \left( Q_{\vc,t} 
   \right)_\alpha \rangle =
   \frac{\partial}{\partial t} \int{\ro u_\alpha}=\int{u_\alpha
   \frac{\partial \ro}{\partial t}} +\int{ \ro \frac{\partial
   u_\alpha}{\partial t}}.
   \ee
 The first term of the right hand side can be calculated again from the
continuity equation. For the second term, we make use of \r{Efelt3} and
\r{lap}. The result is: 
   \be{Ehrenfest} 
   m \frac{\partial^2}{\partial t^2} \langle \left( Q_{\vc,t} 
   \right)_\alpha \rangle =
   \int{\ro {\cal F}^L_\alpha},
   \ee
 where ${\cal F}^L$ has the form of the Lorentz force: 
 $$
   {\cal F}^L_\alpha=e(E_\alpha+\epsilon _{\alpha \beta \gamma} 
   u_\beta B_\gamma).
 $$

It is important to remark that in proving the theorem no special property
of $F$ was used. Therefore, it remains valid even if $F$ is not the
electromagnetic field strength tensor but has any other origin. In that
case $ {\cal F}_\alpha = ( \vj F u )_\alpha = F_{\alpha k} u^k $
appears in the r.h.s.\ of \r{Ehrenfest}.


   \section{\point Nonlinear and nonconservative extensions}


Previously we found that each of the absolute equations \r{lap}, \r{kont},
and \r{wegy}, express independent physical aspects of the quantum system.
This makes it possible to extend quantum mechanics to treat more general
force fields than conservative ones. The extension can be done by
extending or altering Eq.\ \r{wegy}, the equation which describes how
the outer force field acts on the particle, while keeping the mass shell 
condition \r{lap} and the continuity equation \r{kont} untouched. One can
check that, to prove the time independence of $\sa$ by means of the
absolute quantities and equations, only the equations \r{lap} and \r{kont}
are needed. Hence, any alteration in the form of the outer action means no
change in how the structure of the event space is established via $\sa$, 
and the 'unitarity' of the extended system is ensured.
Also, the considerations about states and physical quantities, including
the uncertainty inequalities and a generalized Ehrenfest's theorem, remain
valid.  On the other hand, in replacing \r{wegy} with a new equation, one 
has to check whether the resulting system of equations is self-consistent 
and not too restrictive, and whether the process space of the new system 
is also a Hilbert manifold. 

One possible type of extension is when one keeps the form of \r{wegy} but
replaces $eF$ with a process dependent $\tF$.
For example, one can take $ \tF = \tF(W) $ in the form $ \tF (W)
= d K (W)$, where $ K = K(W) $ is a process dependent covector function.
($K$ has spacetime dependence through $W$, however, in addition, it can
contain explicit spacetime dependence, too.) If $\tF$ is given in this way
then, after a choice of an observer $\vc$, and a spacetime origin $o$, a
quasi Schr\"odinger equation can be obtained from the absolute equations,
if one introduces a `quasi wave function' as
   \be{Psiintegral2} 
   \Psi (p) = R \cdot \exp \left\{ i
   \ftort{\fm}{\fhbar}
   \int_o^p \left[ \left( \ep +
   \ftort{\fegy}{\fm}
   K_0(W) \right) dx^0 + \left( u_\alpha + 
   \ftort{\fegy}{\fm} K_\alpha(W) \right) dx^\alpha \right] \right\}
   \ee
 [cf.\ \r{Psiintegral}]. This quasi Schr\"odinger equation will have the
same form as \r{Sch1}, with $A$ replaced with $K(W)$ or, based on
\r{Psiintegral2}, with $K(\Psi)$. In a coordinate system, it reads
   \be{qSch} 
   i \hbar \left[ \parc\_0 -
   \ftort{\fimag}{\fhbar}
   K\_0 (\Psi) \right] \Psi = -
   \ftort{\fhbarnegyzet}{\fketm}
   \left[ \parc\_\alpha - 
   \ftort{\fimag}{\fhbar}
   K\_\alpha (\Psi) \right] \left[ \parc\_\alpha - 
   \ftort{\fimag}{\fhbar}
   K\_\alpha (\Psi) \right] \Psi.
   \ee 

The well-known, so-called nonlinear Schr\"odinger equation differs from
the usual Schr\"o\-dinger equation in that an additional, $ k | \Psi |^2
\Psi $ term is included. Another known nonlinear modification of the
Schr\"odinger equation is given by Byalinicki-Birula and Mycielski
\cite{BBM}, where the additional term is $ k_1 \ln \left( k_2 | \Psi |
\right) \Psi $. ($k$, $k_1$ and $k_2$ are constants here.) {}From
\r{qSch} we can see that both these modifications are special
cases of a process dependent $ \tF = d K $, with a $\ro$ dependent, timelike
covector function $K$ (recall that timelike covectors have zero space
components in any inertial coordinate system, and an observer invariant
time component). 

The quantum version of the linearly velocity dependent damping can be
formulated by a process dependent $\tF$, too. Let the inertial observer
corresponding to $ \vc \in V(1) $ describe an inertial medium with respect
to which the damping will occur, and we consider $ \tF = - k ( u - \vc ) \ek
\vtau $, with a positive constant $ k \in \vtG / \vtI$. We see that this
$\tF$ is not given in the form $ \tF (W) = d K (W) $. 

For this system, the generalized \r{wegy} has its simplest relative form
[the generalized equations \r{Efelt}, \r{Bfelt}] with respect the inertial
observer $\vc$; we find
   \begin{eqnarray}\l{dE} 
   & \parc_0 u_\alpha = - \frac{k}{m} u_\alpha + \parc_\alpha \ep, &
   \\ \l{dB}  
   & \parc_\alpha u_\beta - \parc_\beta u_\alpha = 0. &
   \end{eqnarray} 
 We remark that if a constant force is present in addition to the damping
force---e.g.,\ a homogeneous gravitational force---, then from $ \tF =
\left[ m g - k ( u - \vc \right] \ek \vtau $, with respect to the inertial
observer $ \vc' = \vc + (m/k) g $, one arrives at the same equations. Thus
all our following considerations will be applicable to this more general 
situation, too.

With respect to the observer $\vc$ (or $\vc'$), the formulas \r{rpont2} 
and \r{Ehrenfest} concerning Ehrenfest's theorem give
   \be{diszEhr}
   \dot{Q}_\alpha = - 
   \ftort{\fk}{\fm}
   \, V_\alpha, \qquad \ddot{Q}_\alpha = -
   \ftort{\fk}{\fm}
   \, \dot{Q}_\alpha,
   \ee
 (for simplicity, the position and velocity expectation values will be
denoted by $Q_\alpha$ and $V_\alpha$ in the following). This shows that
the classical limit of this quantum system is the case of the damping 
which linearly depends on velocity. 

{}From \r{dE} it follows that if \r{dB} holds at a time point then it will
hold later, too. This ensures the consistency of the equations
of the system. Furthermore, \r{dE} and \r{dB} yield that there exists a
real valued spacetime function $S$ such that
   \be{vanS}
   u_\alpha =
   \ftort{\fhbar}{\fm}
   \parc_\alpha S, \qquad \ep =
   \ftort{\fhbar}{\fm}
   \parc_0 S + 
   \ftort{\fk \fspace \fhbar}{\fmnegyzet}
   S.
   \ee
 Hence, by introducing the quasi wave function $ \Psi = R \exp (iS)
$, here we also find that a quasi Schr\"odinger equation is available for
the system. Now this equation reads
   \be{dS}
   i \hbar \parc_0 \Psi = -
   \ftort{\fhbarnegyzet}{\fketm}
   \Laplace \Psi + 
   k \ftort{\fhbar}{\fm}
   S \Psi.
   \ee
 Remarkably, here the extra, nonlinear, `potential-like' term contains the
phase of the wave function, rather than its absolute value. 

In all these three example systems, the found quasi wave function offers 
one way to check that the process space is a Hilbert manifold.


   \section{\point The dissipative system}


In the following we answer some interesting physical questions concerning
the behavior of the above mentioned dissipative system. Namely, we 
are interested
in how damping `stops the particle', whether the spread of a `wave packet'
is also stopped by the damping, and how the kinetic energy of the particle
dissipates. Our findings will be valid for any initial conditions (more
precisely, for all those in which the integrals giving the expectation
values and standard deviations are finite and all surface terms are zero). 
For simplicity, we will perform the calculations for the one space
dimensional version of the system. It is worth observing that the
dimensionful parameters $m$, $\hbar$, and $k$, of the system define the
characteristic length, time and mass scales $ (\hbar / k)\gyok \in
\vtD$, $ m/k \in \vtI $, and $ m \in \vtG $. These distinguished 
units allow us to make the quantities of the system completely 
dimensionless. Dot and prime will mean derivations with respect to
the dimensionless time and space variables, respectively. 

The system is described by the one dimensional form of equations
\r{dE}--\r{dB}, the continuity equation and the mass shell condition. To
obtain a convenient starting point, we take the following observations: In
one space dimension $u$ has only one space component, $u_1$, \r{dB}
becomes trivial, and the mass shell condition [see \r{lap}] can be written
as $ \ep = - \frac{1}{2} u_1^2 + \frac{1}{2} R''/R $ [cf.~\r{sdef}]. Let
us place this expression of $\ep$ into the one dimensional form of \r{dE}:
   $$
   \dot{u}_1 = - u_1 + \ep' = - u_1 + \ftort{\fegy}{\fketto}
   \left\{ - u_1^2 + R'' / R \right\}'.
   $$
 Thus our dissipative system will be governed by this resulting equation,
and by the continuity equation, $ (R^2)\pont + (R^2 u_1)' = 0 $. With the
probability current, $ j_1 = R^2 u_1 $, these two governing equations are
equivalent with the following pair of equations:
   \be{De1}
   \left( R^2 \right)\pont + \left( j\_1 \right)' = 0,
   \ee
   \be{De2}
   \left( j_1 \right)\pont = - j_1 +
   \ftort{\fegy}{\fketto}
   \left\{ R R'' - R'^2 - 2 R^2 u_1^2 \right\}'.
   \ee
 For the following considerations the system of equations \r{De1}--\r{De2}
will be the most practical starting point.

 First let us see whether stationary solutions of these equations 
exist, since if the spread of the `wave packet' stops then it is plausible 
that a 
process will tend to a stationary one, as time tends to infinity. 

Now, for absolute quantities, stationarity means real time independence. 
For a stationary solution, from \r{De1} we find $ \left( j_1 \right)' = 0
$, which means that $j_1$ is both time and space independent. As the space
integral of $j_1 = R^2 u_1$ gives the expectation value of the velocity,
which is assumed to be a finite value, this constant value of $j_1$ must
be zero. The second equation then gives: 
 $$
   \left\{ R R'' - R'^2 \right\}' = 0.
 $$
This equation can be solved, its general solution is:
 $$
   R(x) = c_1 e^{c_0 x} + c_2 e^{-c_0 x},
 $$
 where $c_0$, $c_1$ and $c_2$ are arbitary complex numbers such that $R$
must be real. It is easy to see that for no values of these numbers will
$\ro = R^2$ be normalizable (i.e.,\ having a finite space integral). So
there exists no normalizable stacionary solution of equations \r{De1} and
\r{De2}. 

Next, we determine the time dependence of the position and velocity 
expectation values of the particle. This can be done easily from 
\r{diszEhr}. The result is
   \be{hely}
   Q = Q(0) + V(0) \left( 1 - e^{-t} \right), \qquad V = V(0) e^{-t}.
   \ee

Then we are looking for the asymptotic behavior of the position
expectation value and standard deviation and of the kinetic energy
expectation value. Let $X$ denote the squared standard deviation of the
position, let $Y$ be as in \r{Y}, and let us introduce
 $$
   T := \int \ro \left( u_1 - V \right)^2, \qquad 
   P := \int R'^2.
 $$
 Actually, $T$ and $P$ are nothing else but the two terms on the 
r.h.s.\ of \r{sebszor},
in the one dimensional case. The expectation value of the kinetic energy
can then be given as
   \be{K}
   K = 
   \ftort{\fegy}{\fketto}
   \left( V^2 + T + P \right).
   \ee
{}From \r{De1} and \r{De2} we can deduce the following equations:
   \be{xn1}
   \dot{X}=2Y,
   \ee
   \be{xn2}
   \dot{Y}=-Y+T+P,
   \ee
   \be{xn3}
   (T+P)\dot{}=-2T.
   \ee
The inequalities \r{CSBb} and \r{CSBa} now read
   \be{H1}
   PX \geq 
    \ftort{\fegy}{\fnegy}
   ,
   \ee
   \be{H2}
   TX \geq Y^2.
   \ee
 In our considerations we will several times make use of the facts that
for any functions $f$ and $g$
   \be{tr1}
   \dot{f}+\lambda f = e^{-\lambda t}[e^{\lambda t} f] \, \dot{ },
   \ee
   \be{tr2}
   f \leq g,\,\,\,\, t_1 \leq t_2 \; \; \Rightarrow \; \;
   \int_{t_1}^{t_2} f \leq \int_{t_1}^{t_2} g.
   \ee
The key to find good estimates for $X$ is to rewrite \r{H1} and 
\r{H2} in terms of $X$. In terms of the auxiliary quantity
   \be{Z}
   Z:=(X^2) \, \ddot{ }+(X^2)\dot{ }-3\dot{X}^2,
   \ee
\r{H1} and \r{H2} can be written as simply as
   \be{H1a}
   \dot{Z} \leq 0,
   \ee
   \be{H2a}
   \dot{Z}+2Z \geq 2.
   \ee
 It follows from these inequalities that $Z(t)$ is decreasing and 
converges to a $Z^*$, $ Z^* \geq 1$, as $t$ goes to infinity. Then
   \be{becs1}
   (X^2)\ddot{ }+(X^2)\dot{ }-3\dot{X}^2 \geq Z^*,
   \ee
 or, after omitting a negative term from the left hand side,
   \be{becs2}
   (X^2)\ddot{ }+(X^2)\dot{ } \geq Z^*.
   \ee
 {}From this, after two integrations and using \r{tr1} and \r{tr2}, we 
get
   \be{becs3}
   X^2 \geq Z^*t +C_1-C_2e^{-t},
   \ee
where $C_1$ and $C_2$ are the constants
   $$
   C_1=X^2(0)+(X^2)\dot{ }(0)-Z^*, \qquad C_2=(X^2)\dot{ }(0)-Z^*.
   $$

Now we give an upper bound for $X$. First we prove that $\dot{X}$ is
bounded. It follows from \r{becs3} that there exists a time point $t_1$
such that, for each $t>t_1$, $(X^2)\dot{ }>0$. Since $(X^2)\dot{
}=2X\dot{X}$ and $X \geq 0$, $\dot{X}$ will be positive for each $t>t_1$.
On the other hand, $\dot{X}$ has an upper bound, which can be proven as
follows. $(T+P)\dot{ }=-2T \leq 0$, so $(T+P)(t) \leq (T+P)(0)$ for each
$t>0$. Then
   \be{becs4}
   \dot{Y}+Y \leq (T+P)(0),
   \ee
and applying \r{tr1} and \r{tr2} to this inequality we get
   \be{becs5}
   \ftort{\fegy}{\fketto}
   \dot{X} =Y \leq (T+P)(0)+(Y-T-P)(0)e^{-t}.
   \ee
 Thus $\dot{X}$ is bounded, i.e.,\ there exists a positive number $K_1$ 
such 
that $\dot{X}^2<K_1$ for each $t>t_1$. Using this result we can find 
an upper bound for $Z+3\dot{X}^2$. Let it be $K_2$. Then we can write:
   \be{becs6}
   (X^2)\ddot{ }+(X^2)\dot{ } \leq K_2
   \ee
 for each $t>t_1$. After applying \r{tr1} and \r{tr2} to \r{becs6} we get
the following estimate for $\dot{X}$: 
   \be{becs7}
   0 \leq \dot{X} \leq \frac{K_3}{t\gyok} \qquad (t>t_2),
   \ee
 where $K_3$ and $t_2$ are appropriate constants. We can see that
$\dot{X}$ is not only bounded but it goes to zero when $t$ goes to
infinity. This means that, for every $\epsilon>0$, there exists a $t_3$
such that
   \be{becs8}
   (X^2)\ddot{ }+(X^2)\dot{ } \leq Z^*+\epsilon.
   \ee
 Using \r{tr1} and \r{tr2} again we get:
   \be{becs9}
   X^2 \leq (Z^*+\epsilon)t+C_3-C_4e^{t_3-t},
   \ee
 where $C_3=X^2(t_3)+\dot{X}^2(t_3)-(Z^*+\epsilon)(t_3+1)$ and
$C_4=\dot{X}^2(t_3)-(Z^*+\epsilon)$.

If we compare this to \r{becs3} then we can see how $X$ behaves for 
great values of $t$:
   \be{ered}
   \lim_{t \rightarrow \infty} \frac{X^2(t)}{t} =Z^*.
   \ee
 Hence the standard deviation of the position behaves for great $t$ values
as the function $\sqrt[4]{Z^*} \, t^{1/4}$. This result means that
although the spread of the `wave packet' is slower than in the free
particle case---there $ \sqrt{X} \aranyos t $ asymptotically---, this
spread never stops. 

In parallel, $(X^2)\pont$ tends to $Z^*$ asymptotically. {}From this we
find $ \dot{X} \kb \sqrt{Z^*/4} \, t^{-1/2} $ ($f \kb g$ denotes $ f/g \to
1 $ as $ t \to \infty $), and, from \r{Z}, that $ X \ddot{X} \to 0 $. Thus
$ \ddot{X} / \dot{X} \to 0 $, which, together with \r{xn1} and \r{xn2},
gives $ T + P \kb \sqrt{Z^* / 16} \, t^{-1/2} $. As a consequence,
   \be{ered1}
   K \kb (\sqrt{Z^*}/8) \, t^{-1/2}.
   \ee
 Therefore, we find that the kinetic energy expectation value dissipates
completely, following an asymptotic behavior $t^{-1/2}$.


   \section{\point The Aharonov-Bohm effect}
 

In this section let us turn back to the case in which the outer force 
field is
an electromagnetic field. As we have seen, the absolute formalism---as a
consequence of being gauge ambiguity free---works with the electromagnetic
field strength tensor. This property is an interesting feature from the
aspect of the Aharonov-Bohm effect (see \cite{AB}). In this effect the
particle is excluded from a cylindrical region with an infinite potential 
wall, however, its motion is
influenced by a magnetic field which is nonzero only within this region.
Here we provide an argument that supports that the absolute formalism
may prove to be a good framework to understand the effect purely by means 
of the field strength tensor. Namely, although the probability density is 
zero within the region in question, the following computation will suggest 
that not all
absolute quantities needed for the complete description of the particle
vanish within that domain. Some seem to penetrate into the region,
thus it can be expected that it is these quantities through which the
particle `feels' the action of the inner electromagnetic field. 

Usually, in regions where the electromagnetic field is a finite and smooth
function, $R$ is zero only on a set of measure zero. In the Aharonov-Bohm
effect, however, $R = 0$ in a whole domain. Thus it is a question whether
the other absolute quantities are well defined within this domain. That's
why we apply a limiting procedure. We investigate the set-up with a finite
potential wall, determine the absolute quantities inside and outside the
cylinder, and then we send the potential wall to infinity.  We remark that
in the wave function formalism of quantum mechanics a similar limiting
procedure is needed to derive the boundary conditions the wave function
has to satisfy at an infinite potential wall. We will find that, in the
infinite potential wall limit, some of the absolute quantities have a
nonzero limit inside the cylinder. This suggests that if the absolute
formalism is extended to treat infinite electromagnetic fields, too, the
direct treatment of the Aharonov-Bohm effect will give that the absolute
quantities are well defined inside, too, and some of them are nonzero. 

Let us consider the following set-up. There is an inertial observer $\vc$
and in its space there is an infinitely long cylinder. Let the radius of
the cylinder be $b$. Let our $z$-axis be the axis of the cylinder. If we
fix a space origin on this axis and a direction, perpendicular to the
axis, then every space point can be characterized by its usual cylindrical
coordinates $r$, $\theta$ and $z$. Let the magnetic field be zero outside
the cylinder and a constant value, $B_0$, inside. Let the scalar potential
be zero outside and some function $ \phi (r) $ inside. For our purposes it
will be enough to look for stationary, cylindrically symmetric and $z$
independent solutions of the absolute equations. 

Now, in this case it follows from the continuity equation that $u$ may not
have a radial component, otherwise the outcoming flux of $ j = R^2 u $ on
any cylinder concentric to the original one would not be zero. So $u$ has
only a tangential component which will be denoted by $u_\theta$ and a
$z$-component which will be denoted by $u_z$. The continuity equation then
always holds, and the other equations lead to
   \be{AB1}
   \frac{1}{2}m(u_\theta^2+u_z^2)+e\phi-\frac{\hbar^2}{2m}
   \left( \frac{1}{r}\frac{R'}{R}+\frac{R''}{R}\right)=E,
   \ee
   \be{AB2}
   u_\theta'+\frac{1}{r}u_\theta=\frac{e}{m}B_0,
   \ee
   \be{AB2b}
   u_z'=0
   \ee
 inside the cylinder and
   \be{AB3}
   \frac{1}{2} m (u_\theta^2 + u_z^2) - \frac{\hbar^2}{2m} 
   \left( \frac{1}{r}\frac{R'}{R} + \frac{R''}{R} \right) = E,
   \ee
   \be{AB4}
   u_\theta'+\frac{1}{r}u_\theta=0,
   \ee
   \be{AB4b}
   u_z'=0
   \ee
 outside the cylinder, where $'$ denotes derivatives with respect to $r$,
and the integration constant $E$ is actually the total mechanical energy
of the particle. It follows from \r{AB2b} and \r{AB4b} that $u_z$ is
constant. 

The solution of \r{AB2} is $ u_\theta (r) = \frac{e}{2m} B_0 r + C_1 / r $
(the constant $C_1$ being related to angular momentum, actually), while,
from \r{AB4}, outside the cylinder $ u_\theta (r) = C_2 / r $. Continuity
of $u_\theta$ at $ r = b $ gives $ C_2 = C_1 + \frac{e}{2m} B_0 b^2 $. 

Then, to obtain a convenient way to solve \r{AB1} now we specify the
electromagnetic scalar potential as follows. Let $\phi$ be
$\phi(r)=\phi_0-\frac{eB_0^2}{8m}r^2, \; \; ( 0 \leq r \leq b )$, where
$\phi_0$ is a constant. If $\phi_0$ is sent to infinity then $\phi(r)$
goes to infinity in a uniform way, thus providing the desired infinitely
high potential wall. Eq.\ \r{AB1} then reads
   \be{AB5}
   r^2 R'' + r R' - 
   \ftort{\fketm}{\fhbarnegyzet}
   \left( e\phi_0 - E +
   \ftort{\fegy}{\fketto}
   m u_z^2 + 
   \ftort{\fe}{\fketto}
   B_0 C_1 \right) r^2 R -
   \ftort{\fmnegyzet}{\fhbarnegyzet}
   C_1^2 R = 0.
   \ee
 With $ \kappa = \frac{\sqrt{2m}}{\hbar} \left( e \phi_0 - E + \frac{1}{2}
m u_z^2 + \frac{e}{2} B_0 C_1 \right)\gyok $, this equation is the
modified Bessel equation in the variable $\kappa r$. Its general solution
is $ R (r) = C_3 I_\mu (\kappa r) + C_4 K_\mu (\kappa r) $, where $I_\mu$
and $K_\mu$ are the modified Bessel functions and $ \mu = m C_1 / \hbar$.
Similarly, Eq.\ \r{AB3} can be written as
   \be{AB6}
   r^2 R'' + r R' +
   \ftort{\fketm}{\fhbarnegyzet}
   \left( E -
   \ftort{\fegy}{\fketto}
    m u_z^2 \right) r^2 R -
   \ftort{\fmnegyzet}{\fhbarnegyzet}
   C_2^2 R = 0,
   \ee
 the Bessel equation in the variable $\lambda r$ with $ \lambda =
\frac{\sqrt{2m}}{\hbar} \left( E - \frac{1}{2} m u_z^2 \right)\gyok$.  Its
solution is $ R (r) = C_5 J_\nu (\lambda r) + C_6 Y_\nu (\lambda r) $,
where $ \nu = m C_2 / \hbar $. Matching of $R$ at $r = b$ and the
normalization of $R$ in the variable $r$ determine the constants
$C_3$--$C_6$. 

By using the asymptotic properties of the Bessel functions, it can be
investigated how $R$ tends to zero inside the cylinder in the limit
$\phi_0 \to \infty$. The important issue for our present purpose is to
observe that, in the limit $\phi_0 \to \infty$, $\, u_\theta \,$ (and
probably $u_z$) remains finite and nonzero inside. This makes it plausible
that in the Aharonov-Bohm situation it is the quantity $u$, penetrating
into the cylinder, through which the particle is affected by the inner
magnetic field.


   \section{\point Relativistic considerations}
 

Now let us turn to special relativistic spacetime, and introduce 
absolute quantities and equations concerning the Klein-Gordon 
equation. In
   \be{KG}
   (\hbar^2 D^k D_k+m^2 c^4)\Psi=0
   \ee
 [$ D_k = \parc_k - i (e/\hbar) A_k $, in this section we will work only
with indexed formulas] $\Psi$ is a Lorentz scalar, but has a gauge
dependence similar to that of the nonrelativistic wave function. Using 
again the polar decomposition, $\Psi=Re^{iS}$, let us define
   \be{rudef}
   u_k :=
   \ftort{\fhbar}{\fm}
   \parc_k S -
   \ftort{\fe}{\fm}
   A_k.
   \ee  
 Here the arising consistency relation for $u$ reads
   \be{rel1} 
   \partial_k u_l -\partial_l u_k = -
   \ftort{\fe}{\fm}
   F_{kl}.
   \ee
 With $R$ and $u$, the complex equation \r{KG} will be equivalent to the
following two real ones: 
   \be{rel2} 
   m^2 c^4 R - m^2 g^{kl} u_k u_l R + 
   \hbar^2 g^{kl} \parc_k \parc_l R = 0,
   \ee
   \be{rel3}
   \partial_k (R^2 u^k) =0.
   \ee
 We see that $ j^k := R^2 u^k $ satisfies a continuity equation. As 
indicated before, \r{rel2} is the quantum analogue of the classical mass 
shell relation $ p^k p_k = m^2 c^2$.

It will be interesting to see how the nonrelativistic limit of these 
equations leads to the nonrelativistic absolute equations. 

In general, nonrelativistic limit can be valid only at a neighbourhood of
a spacetime point $p$, and needs an inertial coordinate system with
respect to $ \left| 1 + \frac{u_0(p)}{c^2} \right| \ll 1 $. In the
following we will restrict ourselves to an appropriate neighbourhood of
$p$. To be in conform with the aimed nonrelativistic quantities, we will
have to work with both $\vtD$ and $\vtI$, rather than only with $\vtI$.
Then $ \{ g_{kl} \} = \mbox{diag} [1, - 1/c^2, - 1/c^2, -1/c^2] $, and,
for example, $u_\alpha$ is $ \vtD / \vtI $ valued, and $u_0$ is $ \vtD^\2
/ \vtI^\2 $ valued. 

By introducing 
 $$
   \ep = u_0 + c^2,
 $$
 we can see that $\ep$ and $u_\alpha$ satisfy the nonrelativistic
equations \r{Efelt}, \r{Bfelt}. $ |\ep/c^2| \ll 1 $, and if all terms in
\r{rel3} with higher orders of $1/c^2$ are negligible, then \r{rel3} leads
to the nonrelativistic continuity equation
 $$
   \parc_0 (R^2) + \parc_\alpha ( R^2 u_\alpha ) = 0.
 $$
In \r{rel2}, written as
   \be{rel22}
   \left[ c^2 - c^2 \left( \frac{u_0}{c^2} \right)^2 + 
   u_\alpha u_\alpha \right] R + 
   \frac{\hbar^2}{m^2} \left[ \frac{1}{c^2} \parc_0^2 R - 
   \parc_\alpha \parc_\alpha R \right] = 0,
   \ee
 $ (u_0/c^2)^2 = [ 1 - (\ep/c^2) ]^2 \approx 1 - 2\ep/c^2 $, and if the 
term proportional to $1/c^2$ is negligible again, then we arrive at
   \be{ntomeg}
   \left[ 2 \ep + u_\alpha u_\alpha \right] R + 
   \ftort{\fhbarnegyzet}{\fmnegyzet}
   \parc_\alpha \parc_\alpha R = 0,
   \ee
 which is just the nonrelativistic mass shell condition \r{lap}
[cf.\ \r{sdef}]. 

We can see that the special relativistic $u_0 + c^2$, $u_\alpha$ and $R$
became the nonrelativistic $\ep$, $u_\alpha$ and $R$. 

On a general relativistic spacetime, the straightforward generalization of
the special relativistic Klein-Gordon equation is to replace partial
derivatives with covariant derivatives in \r{KG}. After polar
decomposition, we arrive at quantities and formulas similar to those in the
special relativistic case, again with covariant derivatives instead of
partial derivatives. After some known properties of the covariant
derivative, the equations can be written as
   \be{arel1} 
   \partial_k u_l -\partial_l u_k = -\frac{e}{m} F_{kl},
   \ee
   \be{arel2} 
   m^2 c^4 R - m^2 g^{kl} u_k u_l R + \hbar^2 \frac{1}{\sqrt{-g}} 
   \parc_k \left( \sqrt{-g} g^{kl} \parc_l R \right) = 0,
   \ee
   \be{arel3}
   \frac{1}{\sqrt{-g}} \partial_k 
   \left( \sqrt{-g} g^{kl} R^2 u_l \right) = 0.
   \ee

 In this case we consider the `nonrelativistic + weak gravity' limit of
these equations. In the weak gravity limit, in an appropriate coordinate
system, the matrix of the metric tensor differs from the Minkowskian only
slightly, in the form
 $$
   \{ g_{kl} \} = \mbox{diag} \left[ 1 + \frac{2\phi}{c^2}, 
   - \frac{1}{c^2} \left( 1 - \frac{2\phi}{c^2} \right),
   - \frac{1}{c^2} \left( 1 - \frac{2\phi}{c^2} \right),
   - \frac{1}{c^2} \left( 1 - \frac{2\phi}{c^2} \right) \right],
 $$
 for details, see, e.g.,\ \cite{Wald}. In parallel, $ \left| 1 +
\frac{u_0}{c^2} \right| \ll 1 $ will be needed to hold, at least in a
domain of spacetime, now with respect to this coordinate system. 

In this case we introduce $\ep$ as
 $$
   \ep = u_0 + c^2 + \phi.
 $$
 After such steps as were taken in the special relativistic case, the
calculation gives that, if all terms proportional to higher orders of $
1/c^2 $ are negligible, then \r{arel3} leads to the nonrelativistic
continuity equation, and \r{arel2} to the nonrelativistic mass shell
condition, with $u_\alpha$ and $R$ becoming the corresponding
nonrelativistic quantities. Furthermore, from \r{arel1} we can see that
$\ep$ and $u_\alpha$ satisfy the nonrelativistic \r{Bfelt}, and
   \be{aEfelt} 
   \parc_0 u_\alpha - \parc_\alpha \ep_{\;} =
   \ftort{\fe}{\fm}
   F_{\alpha 0} - \parc_\alpha \phi.
   \ee
 Thus $\phi$ appears as an outer scalar potential acting on the particle,
in addition to the action of the electromagnetic field. This result is
just the expected one: For a classical particle, the presence of a weak
gravity appears as $\phi$ being an outer scalar potential (the Newtonian
gravitational potential) acting on the particle (cf.,\ e.g.,\
\cite{Wald}).


   \section{\point Discussion and outlook}


As we have seen, the presented approach has the following advantages. It
provides a reference frame free and gauge free formulation of quantum
mechanics. The arising quantities are more spacetime friendly and have a
more direct physical interpretation, compared to the wave function. The
three independent physical aspects of the system---namely, the way the
outer field acts on the particle, the mass shell condition, and the
conservation of probability---become transparent, while in the wave
function formalism they remain implicit and hidden. The absolute
framework is applicable to treat nonconservative situations or
`nonlinear' generalizations of quantum mechanics. 

The found results motivate further investigation in diverse directions. 
One of them is to apply the absolute framework for further dissipative
systems. Another one is to study the conjectured connection between the
symplectic structure of the quantum process space and the symplectic
structure of the corresponding classical one. To find applications of the
obtained uncertainty relations is a further arising possibility. 
Besides, by extending the absolute formalism for systems where the given
electromagnetic field contains singularities, a direct way will be
available to study the Aharonov-Bohm situation. This step is important to
check the validity of the indirect treatment presented here. 

To continue establishing the absolute formulation of quantum mechanics,
the future tasks are to extend the presented formalism for multiparticle
quantum mechanics and for particles with spin, in the nonrelativistic
case; and to continue the relativistic cases with the study of the
process space, the states, and the physical quantities, for zero and
nonzero spin particles as well. After these steps the possibility will
open to start building an absolute formalism for quantum field theories.

\bigskip 
 
{\bf Acknowledgements:}
 The authors wish to thank Tam\'as Matolcsi for valuable discussions,
Izumi Tsutsui for valuable remarks and K\'aroly Banicz for useful
suggestions. T. F. is grateful to the Japan Society for the Promotion of
Science for financial support and to IPNS, KEK Tanashi Branch, Tokyo,
Japan, for their kind hospitality. 
 



\begin{thebibliography}{99} 

 

\bibitem{FesVil} 
H. Feshbach and F. Villars, {\it Rev. Mod. Phys.} {\bf 30} (1958), 24.

\bibitem{FT2}
T. F\"ul\"op, ``Relativistic quantum mechanics on the SL(2, R)
space--time,'' {\it J. Math. Phys.} {\bf 38} No. 2 (1997), 611-621. 

\bibitem{ASpot} 
A. Ashtekar and J. Stachel, ``Conceptual problems of quantum 
gravity,'' Birkh\"auser, Boston, 1991.

\bibitem{Kuc}
K. V. Kucha\v{r}, ``Time and interpretations of quantum
gravity,'' {\it in} ``Proceedings of the 4th Canadian conference on
general relativity and relativistic astrophysics,'' (G.
Kunstatter, D. Vincent and J. Williams, Eds.), World Scientific,
Singapore, 1992.

\bibitem{Ish} 
C. J. Isham, ``Canonical quantum gravity and the problem of time,'' {\it
in} ``Integrable systems, quantum groups and quantum field theory,'' (L.A.
Ibart and M.A. Rodrigues, Eds.), Kluwer, Dordrecht, 1992. 

\bibitem{KT} 
T. Kashiwa and N. Tanimura, ``Gauge independence in terms of the  
functional integral,'' {\it Phys. Rev.} {\bf D56} (1997), 2281-90.  
 
\bibitem{Weyl}
H. Weyl, ``Space-Time-Matter'', Dover publ. 1922.
 
\bibitem{MT3} 
T. Matolcsi, ``Spacetime without reference frames,'' Aka\-d\'e\-mi\-ai
Ki\-a\-d\'o, Budapest, 1993. 

\bibitem{MT1} 
T. Matolcsi, ``A concept of mathematical physics, Models for spacetime,''
A\-ka\-d\'e\-mi\-ai Ki\-a\-d\'o, Budapest, 1984. 

\bibitem{Sou} 
J.M. Souriau, ``Structure des syst\`emes dynamique,'' Dunod, Paris, 
1970.

\bibitem{MT2} 
T. Matolcsi, ``A concept of mathematical physics, Models in mechanics,''
A\-ka\-d\'e\-mi\-ai Ki\-a\-d\'o, Budapest, 1986. 

\bibitem{Mad} 
E. Ma\-de\-lung, {\it Z. Phys.} {\bf 40} (1926), 322.

\bibitem{Tak} 
T. Takabayashi, {\it Suppl. Progr. Theor. Phys.} {\bf 4} (1957), 1.
 
\bibitem{Jan}
L. J\'a\-nossy, {\it Ac\-ta Phys. Hung. Tom.} {\bf XVI} (1963), 37. 

\bibitem{BH}
D. Bohm and B. J. Hiley, ``The Undivided Universe: an ontological
interpretation of quantum theory,'' Routledge, London and New York, 1993.

\bibitem{KR} 
J. Kijowski, G. Rudolph, {\it Phys. Rev.} {\bf D31} (1985), 856; {\it
Nucl.  Phys.} {\bf B325} (1989), 211; {\it Lett. Math. Phys.} {\bf 29}
(1993), 103. 
 
\bibitem{AS}
A. Ashtekar and T. A. Schilling, ``Geometrical formulation of quantum
mechanics,'' preprint CGPG 97/6-1; LANL xxx archive server, E-print No. 
gr-qc/9706069. 
 
\bibitem{BBM}
I. Bialynicki-Birula and J. Mycielski, ``Nonlinear wave mechanics,'' 
{\it Ann. Phys. (N.Y.)} {\bf 100} (1976), 62-93.

\bibitem{W}
S. Weinberg, ``Testing quantum mechanics,'' {\it Ann. Phys. (N.Y.)} {\bf
194} (1989), 336-386. 

\bibitem{AB} 
Y. Aharonov and D. Bohm, {\it Phys. Rev.} {\bf 115} (1959), 485-491. 
 
\bibitem{Mac}
G. Mackey, ``The mathematical foundation of quantum mechanics,''
Benjamin, Inc.,\ New York, 1963. 

\bibitem{Seg}
I. Segal, ``Mathematical problems of relativistic physics,'' Am. Math.
Soc. Prov. Rhode Island, 1963. 

\bibitem{Gud}
S. Gudder, ``Convex structures and operational quantum mechanics,'' 
{\it Comm. Math. Phys.} {\bf 29} (1973), 249-264. 

\bibitem{Wald}
R.M. Wald, ``General relativity,'' Univ. of Chicago Press, Chicago,
1984. 

   \end{thebibliography}
\end{document}